\documentclass[aps,prd,pdflatex,reprint,10pt,superscriptaddress,floatfix,nofootinbib]{revtex4-2}

\usepackage{float}
\usepackage{amsmath}
\usepackage[T1]{fontenc}
\usepackage[utf8]{inputenc}
\usepackage{lmodern}
\usepackage{amssymb}
\usepackage{natbib}
\usepackage{graphicx}
\usepackage{hyperref}
\usepackage{epstopdf}
\usepackage{pifont}
\usepackage{textgreek}
\usepackage[normalem]{ulem}
\usepackage{placeins}
\usepackage{tikz}
\usepackage{physics}
\usepackage{orcidlink}
\usepackage{array}
\usepackage{booktabs,multirow,siunitx}

\usetikzlibrary{calc}
\graphicspath{{pdf/}}

\hypersetup{
 colorlinks=true,
 linkcolor=blue,
 citecolor=blue,
 filecolor=blue,
 urlcolor=blue,
 breaklinks=true
}

\RequirePackage{color}

\newcommand{\placeholderfig}[2][0.92\linewidth]{%
  \fbox{%
    \begin{minipage}[c][5.0cm][c]{#1}
      \centering
      \textbf{#2}
    \end{minipage}%
  }%
}

\newcommand{\includefileorplaceholder}[2]{%
  \IfFileExists{#1}
  {\includegraphics[width=\linewidth]{#1}}
  {\placeholderfig{#2\\[0.4em]\texttt{\detokenize{#1}}}}%
}

\begin{document}

\date{\today}

\title{Topological control of third-harmonic generation in a mesoscopic quantum ring with spiral dislocation}

\author{Carlos Magno O. Pereira}
\email{carlos.mop@discente.ufma.br}
\affiliation{
Departamento de F\'isica,
Universidade Federal do Maranh\~ao,
65085-580 S\~ao Lu\'is, MA, Brazil}

\author{Denise Assafr\~ao}
\email{denise.lima@ufes.br}
\affiliation{
Departamento de F\'isica,
Universidade Federal do Esp\'irito Santo,
Vit\'oria, ES, Brazil}

\author{Edilberto O. Silva\orcidlink{0000-0002-0297-5747}}
\email[Edilberto O. Silva - ]{edilberto.silva@ufma.br}
\affiliation{
Departamento de F\'isica,
Universidade Federal do Maranh\~ao,
65085-580 S\~ao Lu\'is, MA, Brazil}

\begin{abstract}
We investigate the nonlinear optical response of a two-dimensional mesoscopic quantum ring subjected to a spiral dislocation, with emphasis on the third-harmonic generation (THG). The topological defect is modeled through a torsion-induced deformation of space that modifies the effective metric without introducing curvature. By combining the minimal-coupling prescription in curved space with a radial ring confinement and a perpendicular magnetic field, we derive the effective radial Schr\"odinger problem, obtain the bound states, and evaluate the nonlinear susceptibilities within the electric-dipole approximation. We show that the axial symmetry of the topologically deformed ring preserves the selection rule $\Delta m=\pm 1$ and therefore suppresses second-harmonic generation in the dipole approximation, whereas THG remains allowed through multistep transition chains. The study is further expanded through three complementary analyses that can be implemented without changing the Hamiltonian: a dephasing-controlled study of spectral resolution and a channel-resolved decomposition of the THG amplitude. Together, these results and extensions establish spiral dislocation as a robust geometric knob for tuning nonlinear optical activity in mesoscopic ring-shaped nanostructures.
\end{abstract}


\maketitle

\section{Introduction}

Quantum rings occupy a distinctive place among low-dimensional semiconductor structures because their doubly connected geometry allows confinement, phase coherence, and magnetic flux to interplay on equal footing. Since the seminal proposal of Aharonov and Bohm, ring geometries have served as paradigmatic platforms for observing phase-sensitive quantum phenomena, including Aharonov--Bohm oscillations, persistent currents, and field-driven changes in orbital angular momentum \cite{Aharonov1959,Lorke2000,Fuhrer2001,RevModPhys.57.339,PRL.1985.55.1610,PRL.1993.70.2020,PhysRevB.72.125348,PhysRevB.74.125426,TanInkson1999,PRB.2006.73.195310,PhysRevB.77.205303,EPJB.2006.53.99}. In semiconductor implementations, these effects can be combined with strong lateral confinement, making quantum rings especially attractive for optoelectronic and spectroscopic applications.

In addition to their magnetic and transport signatures, quantum rings exhibit a rich optical response that is highly sensitive to geometry, external fields, and confinement engineering. Linear and nonlinear absorption, refractive-index changes, and harmonic generation have all been shown to depend strongly on the details of the ring potential, the applied magnetic flux, the impurity configuration, and the structural asymmetry \cite{Liang2011,Salehani2023,Chang2023,Lima2023,Bejan2025,Duque2010,AdP.2012.524.327,rashba,ringparallel,OM.2019.91.309,SR.2019.9.1427,SM.2013.58.94,CTP.2024.76.105701,QR.2024.6.677}. In particular, third-harmonic generation (THG) is an especially useful probe because it combines strong spectral selectivity with direct sensitivity to the underlying level spacings and dipole matrix elements. This makes THG a natural observable for identifying how topology and geometry reshape the multilevel structure of mesoscopic rings.

From a modeling perspective, a key point is that the nonlinear response of ring-confined carriers is governed by the same ingredients that control the intraband and intersubband spectra: confinement geometry, field-driven level (anti)crossings, and symmetry-imposed selection rules. In axially symmetric rings, the electric-dipole operator couples angular momenta according to $\Delta m=\pm 1$, which can strongly restrict even-order processes in the dipole approximation while still allowing odd-order generation through multistep pathways. This interplay between symmetry, multilevel structure and dephasing underlies a wide range of nonlinear-optical predictions in quantum-confined heterostructures and provides the natural framework for discussing SHG suppression and THG enhancement in the present system \cite{ahn,rosencher,Duque2010,LI2017375,doi:10.1142/S0217979217500096,AdP.2012.524.327}.

A complementary route to controlling quantum-ring optics is to embed the carrier dynamics in a nontrivial geometric background produced by topological defects. In the geometric theory of defects, dislocations and disclinations are described in terms of curvature and torsion within a Riemann--Cartan framework \cite{Katanaev1992,Katanaev2005,FP.1980.10.151}. Spiral dislocations are particularly interesting because they introduce torsion while preserving flat curvature, thereby mixing radial and azimuthal coordinates without destroying the ring topology itself \cite{Valanis2005,EPL.1999.45.279,AofP.2014.346.51,PhysRevA.87.012130,PLA.2015.379.2110,PLA.2016.380.3847}. Over the last few years, spiral-dislocation backgrounds have been explored in connection with harmonic oscillators, Landau quantization, quantum revivals, and confined ring states \cite{Maia2018,daSilva2020,Hassanabadi2026,AoP.2020.419.168229,EPJC.2019.79.551,EPJC.2025.85.34,EPJP.2022.137.589,EPJA.2021.57.192,IJP.2024.98.4827}. Very recently, the same geometric setting was shown to alter optical absorption, refractive-index changes, and photoionization cross-sections in quantum rings, strongly suggesting that harmonic generation should also be controllable via the defect strength \cite{Hassanabadi2026,AdP.2011.523.898}.

The purpose of the present work is to build that nonlinear-optical extension. We consider a charged spinless carrier confined to a two-dimensional mesoscopic quantum ring with harmonic radial confinement, subjected simultaneously to a perpendicular magnetic field and a spiral dislocation. We first derive the corresponding effective Hamiltonian and radial Schr\"odinger-like equation in the torsion-deformed background. Then we analyze the low-lying spectrum, radial probability densities, and dipole-allowed transitions, and use them to construct the second- and third-order susceptibilities. A central point of the paper is that the defect modifies the radial structure and the transition amplitudes while preserving axial symmetry, so that second-harmonic generation (SHG) remains forbidden in the electric-dipole approximation, whereas THG survives through allowed multistep channels \cite{ahn,rosencher,LI2017375,doi:10.1142/S0217979217500096}.

Furthermore, we expand our study through three complementary analyses: (i) a dephasing-dependent analysis of the THG linewidths and spectral resolution, (ii) 3D waterfall spectra that expose the control exerted by spiral dislocation and magnetic field, and (iii) a decomposition of the total THG signal into symmetry-allowed transition channels and their interference terms.

\section{Theoretical model and effective Hamiltonian}
\label{sec:model}

We consider a two-dimensional quantum ring in the presence of a spiral dislocation and a uniform magnetic field $\mathbf{B}=B\hat{z}$. In polar coordinates $(r,\varphi)$, the dislocation is described by the effective metric \cite{Valanis2005,daSilva2020,Hassanabadi2026}
\begin{equation}
 ds^2 = dr^2 + 2\beta\,dr\,d\varphi + (\beta^2+r^2)\,d\varphi^2,
 \label{eq:metric}
\end{equation}
where $\beta$ measures the dislocation strength. The metric tensor is
\begin{equation}
 g_{ij}=\begin{pmatrix}1 & \beta\\ \beta & \beta^2+r^2\end{pmatrix},
 \qquad \det(g_{ij})=g=r^2,
 \label{eq:metric_tensor}
\end{equation}
and the corresponding inverse metric is
\begin{equation}
 g^{ij}=\frac{1}{r^2}\begin{pmatrix}\beta^2+r^2 & -\beta\\ -\beta & 1\end{pmatrix}.
 \label{eq:inverse_metric}
\end{equation}
Therefore, the dislocation changes the local coupling between radial and angular motion without introducing Gaussian curvature.

The carrier is modeled within the effective-mass approximation by the stationary Schr\"odinger equation in curved space \cite{PhysRev.152.683,vonRoos,Morrow,PhysRevA.66.042116,PRA.1999.60.4318,dewitt1957dynamical,dacosta},
\begin{equation}
\begin{split}
 E\Psi ={}& -\frac{\hbar^2}{2\mu}\frac{1}{\sqrt{g}}
 \left(\partial_k-i\frac{e}{\hbar}A_k\right)
 \left[\sqrt{g}\,g^{kj}
 \left(\partial_j-i\frac{e}{\hbar}A_j\right)\Psi\right] \\
 &+V_{\mathrm{conf}}(r)\Psi .
\end{split}
 \label{eq:schrodinger_curved}
\end{equation}
where $\mu$ is the effective mass, $e>0$ denotes the magnitude of the elementary charge, and $V_{\mathrm{conf}}(r)$ models the ring confinement. We adopt the displaced parabolic potential
\begin{equation}
 V_{\mathrm{conf}}(r)=\frac{1}{2}\mu\omega_0^2(r-r_0)^2,
 \label{eq:confinement}
\end{equation}
whose minimum at $r=r_0$ sets the mean ring radius. For a perpendicular magnetic field, the symmetric gauge is used is $A_{\hat{\varphi}}=Br/2$. 

Following the standard convention commonly used in the quantum-ring literature, particularly in the exactly soluble models of Tan and Inkson~\cite{TanInkson1996,TanInkson1999}, we take $e>0$ as the magnitude of the elementary charge and regard $B$ as a signed magnetic-field parameter. With this convention, the electron magnetic coupling is written in the compact form $\partial_j-i(e/\hbar)A_j$, with $A_j$ defined above, and the cyclotron frequency is
\begin{equation}
 \omega_c=\frac{eB}{\mu}.
 \label{eq:cyclotron_frequency}
\end{equation}
Thus, the sign associated with the electronic charge is encoded in the orientation of the magnetic-field parameter $B$, while $e$ is kept positive throughout the analytical expressions and numerical implementation.

Using the separable ansatz
\begin{equation}
 \Psi(r,\varphi)=e^{im\varphi}f(r),
 \label{eq:ansatz}
\end{equation}
with integer angular quantum number $m$, Eq.~\eqref{eq:schrodinger_curved} reduces to the radial equation
\begin{equation}
 A(r)f''(r)+B(r)f'(r)+Q(r)f(r)=0,
 \label{eq:radial_eq}
\end{equation}
where
\begin{align}
 A(r) &= 1+\frac{\beta^2}{r^2},
 \label{eq:Acoef}\\[2mm]
 B(r) &= \frac{1}{r}-\frac{\beta^2}{r^3}
 -\frac{2i\beta m}{r^2}+i\frac{\mu\omega_c\beta}{\hbar},
 \label{eq:Bcoef}\\[2mm]
 Q(r) &= -\frac{m^2}{r^2}+\frac{im\beta}{r^3}
 -\frac{\mu^2\omega_c^2r^2}{4\hbar^2}
 +i\frac{\mu\omega_c\beta}{2\hbar r}\notag\\
 &\quad +\frac{\mu\omega_cm}{\hbar}
 +\frac{2\mu E}{\hbar^2}
 -\frac{2\mu}{\hbar^2}V_{\mathrm{conf}}(r).
 \label{eq:Qcoef}
\end{align}
The mixed radial-angular term in the metric produces the imaginary contributions in Eqs.~\eqref{eq:Bcoef} and \eqref{eq:Qcoef}. These terms originate from the combined action of torsion and magnetic coupling; they do not imply a non-Hermitian spectrum, because the full operator in Eq.~\eqref{eq:schrodinger_curved} remains Hermitian with respect to the curved-space inner product.

\begin{figure}[tbhp]
\centering
\includegraphics[width=\columnwidth]{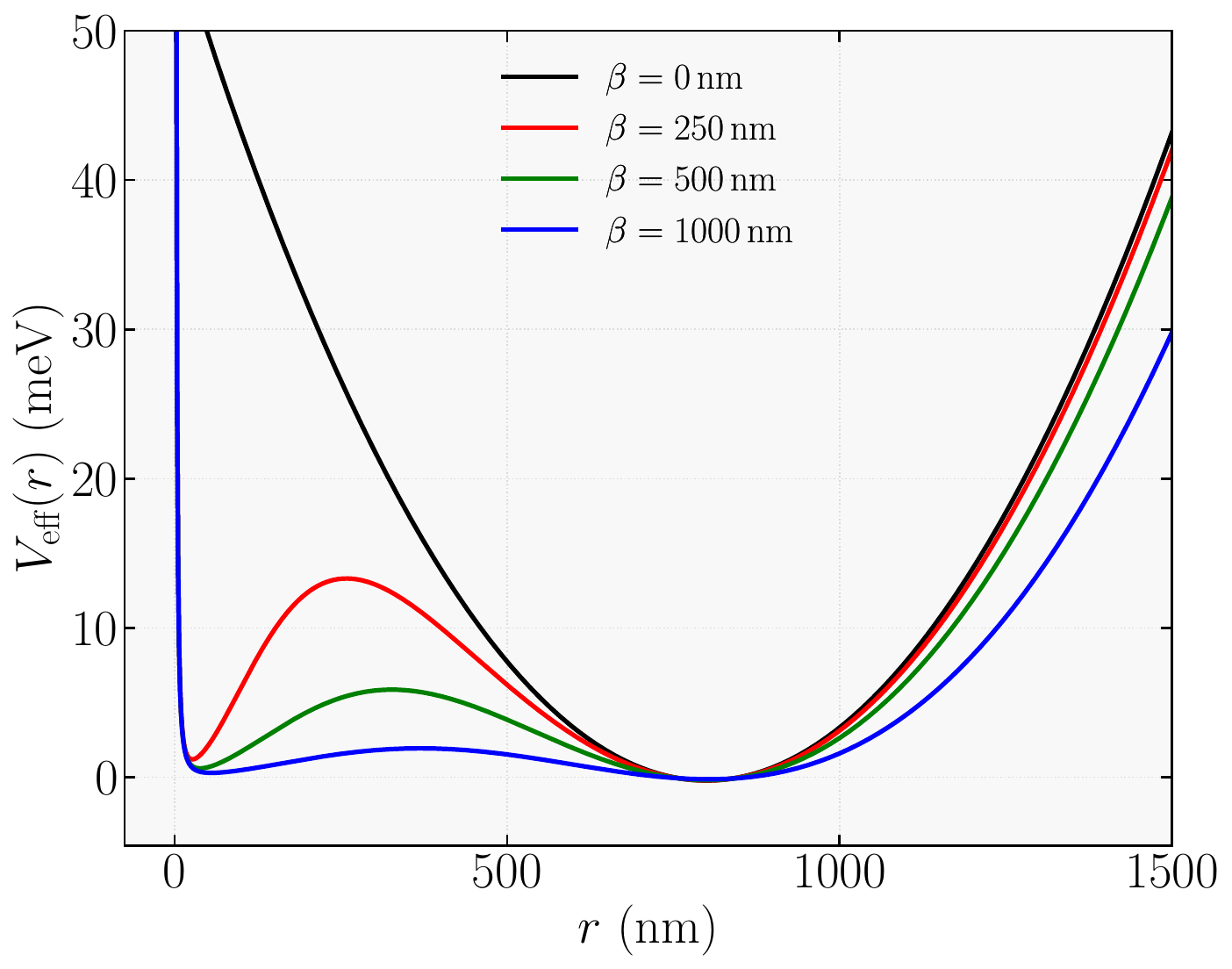}
\caption{\footnotesize (Color online) Real part of the effective radial potential, $\mathrm{Re}[V_{\mathrm{eff}}(r;E)]$, for different values of the spiral-dislocation parameter $\beta$ at zero magnetic field ($B = 0$). The potential is evaluated at the converged reference energy used for the displayed low-lying branch. The confinement parameters are fixed to $r_0 = 800\,\mathrm{nm}$ and $\hbar\omega_0 = 0.449\,\mathrm{meV}$.}
\label{fig:veff}
\end{figure}

To express Eq.~\eqref{eq:radial_eq} into a one-dimensional Schr\"odinger-like form, we remove the first derivative through the transformation
\begin{equation}
 f(r)=\chi(r)\exp\left[-\frac{1}{2}\int^r\frac{B(\rho)}{A(\rho)}d\rho\right].
 \label{eq:chi_transform}
\end{equation}
This yields
\begin{equation}
 -\frac{\hbar^2}{2\mu}\chi''(r)+V_{\mathrm{eff}}(r;E)\chi(r)=0,
 \label{eq:chi_eq}
\end{equation}
with the energy-dependent effective potential
\begin{equation}
\begin{split}
 V_{\mathrm{eff}}(r;E)={}&-\frac{\hbar^2}{2\mu}
 \Bigg[
 \frac{Q(r)}{A(r)}
 -\frac{1}{2}\frac{d}{dr}\left(\frac{B(r)}{A(r)}\right) \\
 &\hspace{2.5cm}
 -\frac{1}{4}\left(\frac{B(r)}{A(r)}\right)^2
 \Bigg].
\end{split}
 \label{eq:veff}
\end{equation}
The structure of the effective potential in Eq.~\eqref{eq:veff} provides a direct way to visualize the geometric action of the spiral dislocation on the radial confinement. Since $V_{\mathrm{eff}}(r;E)$ is energy dependent and may contain complex intermediate contributions after the first-derivative removal, the physically relevant diagnostic is obtained by evaluating its real part at the converged reference energy of the state under consideration. This quantity is displayed in Fig.~\ref{fig:veff} for representative values of the dislocation parameter $\beta$. The figure shows that the defect does not merely shift the radial confinement uniformly; instead, it reshapes the effective radial landscape and can generate an additional inner-well structure between the origin and the mean ring radius. This deformation anticipates the redistribution of the radial probability densities and the transition matrix elements discussed below.

The explicit energy dependence in Eq.~\eqref{eq:veff} requires a self-consistent eigenvalue procedure. In practice, one fixes a trial energy, constructs $V_{\mathrm{eff}}$, solves Eq.~\eqref{eq:chi_eq} numerically, and iterates until the input and output energies converge.

\section{Nonlinear optical formalism}
\label{sec:nonlinear}
The study of the nonlinear optical response begins with the consideration that a monochromatic optical radiation field $\mathbf{E}(t)$ with frequency $\omega$, linearly polarized along the $x$-direction, is applied to the system, which can be expressed by:
\begin{equation}
	\mathbf{E}(t) = \tilde{\mathbf{E}}e^{i\omega t} + \tilde{\mathbf{E}}e^{-i\omega t}.
\end{equation}
The nonlinear optical properties of the system are determined by means of the density matrix operator formalism $\hat{\rho}$ to demonstrate the suppression of the Second Harmonic Generation (SHG) and to calculate the Third Harmonic Generation (THG) characteristics in the topological effects of the spiral dislocation in the quantum ring. For a one-electron system, the time evolution of the matrix elements of $\hat{\rho}$ is described by the von Neumann equation \cite{CAP.2011.11.181,ahn,rosencher,Duque2010,LI2017375,doi:10.1142/S0217979217500096}:
\begin{equation}
 \frac{\partial\hat{\rho}_{ij}}{\partial t}=
 \frac{1}{i\hbar}
 \left[\hat{H}_{0}-\hat{M}\cdot\mathbf{E}(t),\hat{\rho}\right]_{ij} -\widetilde{\Gamma}_{ij}
 \left(\hat{\rho}-\hat{\rho}^{(0)}\right)_{ij} .
\end{equation}
where $\hat{\rho}$ is the density matrix of the one-electron system, $\hat{\rho}^{(0)}$ is the unperturbed density matrix operator, $\hat{H}_{0}$ is the Hamiltonian of the system in the absence of the electromagnetic field, $\hat{M}\cdot\mathbf{E}(t) = -e\hat{\mathbf{r}}E(t)$ represents the perturbative electric dipole interaction contribution, and $\widetilde{\Gamma}_{ij}$ is the phenomenological relaxation rate. In the adopted model, considering linearly polarized light, the dipole operator is specifically given by $\hat{x} = r\cos\varphi$.

To solve the evolution equation, we expand the density matrix in a perturbation series:
\begin{equation}
	\hat{\rho}(t) = \sum_{n=0}^{\infty}\hat{\rho}^{(n)}.
\end{equation}

Substituting this expansion into the time evolution equation, we obtain the following iterative relation:
\begin{equation}
\begin{split}
 \frac{\partial\hat{\rho}_{ij}^{(n+1)}}{\partial t}={}&
 \frac{1}{i\hbar}
 \left(\left[\hat{H}_0,\hat{\rho}^{(n+1)}\right]_{ij}
 -i\hbar\widetilde{\Gamma}_{ij}\hat{\rho}_{ij}^{(n+1)}\right) \\
 &-\frac{1}{i\hbar}
 \left[e\hat{\mathbf{r}},\hat{\rho}^{(n)}\right]_{ij}E(t) .
\end{split}
\end{equation}
Considering the interaction with the field $E(t)$, the resulting macroscopic polarization induced in the system can be expressed as  \cite{PBCM.2023.665.415042,ahn,rosencher,Duque2010}:
\begin{equation}
\begin{split}
 P(t)={}&\varepsilon_{0}\chi_{\omega}^{(1)}\tilde{E}e^{i\omega t}
 +\varepsilon_{0}\chi_{0}^{(2)}\tilde{E}^{2} \\
 &+\varepsilon_{0}\chi_{2\omega}^{(2)}\tilde{E}^{2}e^{2i\omega t}
 +\varepsilon_{0}\chi_{\omega}^{(3)}\tilde{E}^{3}e^{i\omega t} \\
 &+\varepsilon_{0}\chi_{3\omega}^{(3)}\tilde{E}^{3}e^{3i\omega t}+\cdots .
\end{split}
\end{equation}
where $\varepsilon_{0}$ is the vacuum permittivity; and the terms $\chi_{\omega}^{(1)}$, $\chi_{0}^{(2)}$, $\chi_{2\omega}^{(2)}$, $\chi_{\omega}^{(3)}$, and $\chi_{3\omega}^{(3)}$ are the linear, optical rectification, second harmonic generation (SHG), third-order, and third harmonic generation (THG) susceptibilities, respectively. 

\subsection{Dipole matrix elements and selection rules}
\label{sec:selection_rules}

The eigenstates of the ring are denoted by $\ket{n,m}$, where $n$ labels the radial mode and $m$ the azimuthal angular momentum. Their coordinate representation is written as
\begin{equation}
 \Psi_{n,m}(r,\varphi)=\frac{1}{\sqrt{2\pi}}R_{n,m}(r)e^{im\varphi}.
 \label{eq:eigenstate_form}
\end{equation}
For linearly polarized light along the $x$ direction, the electric-dipole operator is
\begin{equation}
 \hat{x}=r\cos\varphi.
 \label{eq:dipole_operator}
\end{equation}
The corresponding dipole matrix element between states $\ket{i}$ and $\ket{j}$ is
\begin{equation}
 M_{ij}=e\braket{i|r\cos\varphi|j}.
 \label{eq:dipole_matrix}
\end{equation}
Using Eq.~\eqref{eq:eigenstate_form}, the angular integral immediately gives
\begin{equation}
 \Delta m = \pm 1,
 \label{eq:selection_rule}
\end{equation}
and diagonal matrix elements vanish:
\begin{equation}
 M_{ii}=0.
 \label{eq:no_diagonal_dipole}
\end{equation}
These two properties strongly constrain the nonlinear response.

\subsection{Why second-harmonic generation is absent}
\label{sec:shg_absence}

Within the standard sum-over-states density-matrix formalism, the second-order susceptibility can be written in an energy-denominator notation as
\begin{equation}
 \chi^{(2)}(2\omega)=\frac{N}{\varepsilon_0}
 \sum_{a,b}\frac{M_{ga}M_{ab}M_{bg}}{D_{ab}^{(2)}(\omega)}
 +\mathrm{perm.},
 \label{eq:chi2}
\end{equation}
with
\begin{equation}
\begin{split}
 D_{ab}^{(2)}(\omega)={}&
 (\Delta E_{ag}-2\hbar\omega-i\Gamma_{ag}) \\
 &\times(\Delta E_{bg}-\hbar\omega-i\Gamma_{bg}) .
\end{split}
 \label{eq:Dab2}
\end{equation}
Here $\ket{g}$ is the initial state, $\Delta E_{ij}=E_i-E_j$, and $N$ is the effective carrier density entering the macroscopic polarization, which should not be confused with the radial quantum number $n$. The microscopic dephasing rate is denoted by $\widetilde{\Gamma}_{ij}=1/T_{2,ij}$, whereas $\Gamma_{ij}=\hbar\widetilde{\Gamma}_{ij}=\hbar/T_{2,ij}$ denotes the corresponding dephasing broadening in energy units. This energy-denominator convention is used throughout the numerical spectra, where both $\hbar\omega$ and $\Gamma$ are expressed in meV.

A nonzero term in Eq.~\eqref{eq:chi2} requires the three dipole factors to satisfy Eq.~\eqref{eq:selection_rule} simultaneously:
\begin{align}
 M_{ga}\neq 0 &\Rightarrow m_a=m_g\pm 1,
 \label{eq:shg_cond1}\\
 M_{ab}\neq 0 &\Rightarrow m_b=m_a\pm 1,
 \label{eq:shg_cond2}\\
 M_{bg}\neq 0 &\Rightarrow m_b=m_g\pm 1.
 \label{eq:shg_cond3}
\end{align}
These conditions cannot all be fulfilled without violating the dipole rule or generating a forbidden diagonal element. Hence,
\begin{equation}
 \chi^{(2)}(2\omega)=0.
 \label{eq:chi2_zero}
\end{equation}
This result is robust with respect to the spiral dislocation: although $\beta$ reshapes the radial wave functions and lifts the degeneracy between states with opposite $m$ through the magnetic coupling, the Hamiltonian remains independent of the azimuthal coordinate. Therefore, $m$ is still a good quantum number and the suppression of SHG persists for all values of $\beta$ within the electric-dipole approximation.
\begin{figure*}[tbhp]
\centering
\includegraphics[width=0.8\linewidth]{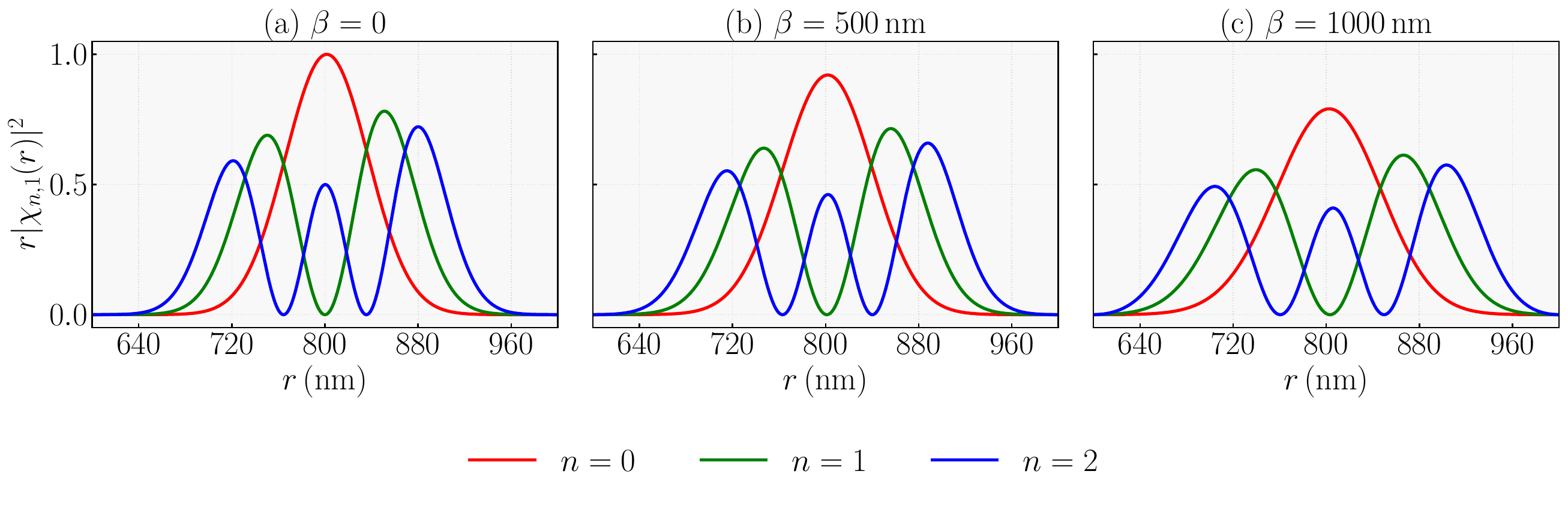}
\caption{\footnotesize (Color online) Normalized radial probability densities $r|\chi_{n,1}(r)|^2$ for the low-lying states ($n = 0, 1, 2$) for different spiral dislocation strengths $\beta$ and zero magnetic field ($B = 0$).}
\label{fig:prob_density}
\end{figure*}

\subsection{Third-harmonic generation}
\label{sec:thg_formalism}

The third-order susceptibility associated with emission at $3\omega$ is written as
\begin{equation}
 \chi^{(3)}(3\omega)=\frac{N}{\varepsilon_0}
 \sum_{a,b,c}
 \frac{\mathcal{M}_{gabc}}
 {D_{abc}^{(3)}(\omega)}
 +\mathrm{perm.},
 \label{eq:chi3}
\end{equation}
where
\begin{equation}
 \mathcal{M}_{gabc}=M_{ga}M_{ab}M_{bc}M_{cg},
 \label{eq:Mgabc}
\end{equation}
and
\begin{equation}
\begin{split}
 D_{abc}^{(3)}(\omega)={}&
 (\Delta E_{ag}-\hbar\omega-i\Gamma_{ag}) \\
 &\times(\Delta E_{bg}-2\hbar\omega-i\Gamma_{bg}) \\
 &\times(\Delta E_{cg}-3\hbar\omega-i\Gamma_{cg}) .
\end{split}
 \label{eq:Dabc}
\end{equation}
In contrast with SHG, the product in Eq.~\eqref{eq:Mgabc} contains four dipole matrix elements, so one can build closed transition chains in which each individual step obeys Eq.~\eqref{eq:selection_rule}. Two representative families starting from $\ket{0,0}$ are
\begin{align}
 \ket{0,0} &\rightarrow \ket{n_1,1}
 \rightarrow \ket{n_2,0} \notag\\
 &\rightarrow \ket{n_3,1}
 \rightarrow \ket{0,0},
 \label{eq:pathA}\\
 \ket{0,0} &\rightarrow \ket{n_1,1}
 \rightarrow \ket{n_2,2} \notag\\
 &\rightarrow \ket{n_3,1}
 \rightarrow \ket{0,0}.
 \label{eq:pathB}
\end{align}
The corresponding symmetry-related channels with negative angular momenta are
\begin{align}
 0 &\rightarrow -1 \rightarrow 0 \rightarrow -1 \rightarrow 0,
 \label{eq:pathAneg}\\
 0 &\rightarrow -1 \rightarrow -2 \rightarrow -1 \rightarrow 0.
 \label{eq:pathBneg}
\end{align}
For a single path $P$, it is convenient to define the partial amplitude
\begin{equation}
 \mathcal{A}_P(\omega)=\frac{N}{\varepsilon_0}
 \frac{M_{ga}M_{ab}M_{bc}M_{cg}}{D_P^{(3)}(\omega)},
 \label{eq:path_amplitude}
\end{equation}
where, for the path $P:g\to a\to b\to c\to g$,
\begin{equation}
\begin{split}
 D_P^{(3)}(\omega)={}&
 (\Delta E_{ag}-\hbar\omega-i\Gamma) \\
 &\times(\Delta E_{bg}-2\hbar\omega-i\Gamma) \\
 &\times(\Delta E_{cg}-3\hbar\omega-i\Gamma) .
\end{split}
 \label{eq:DP3}
\end{equation}
The total susceptibility is then obtained by coherent summation over all allowed paths:
\begin{equation}
 \chi^{(3)}(3\omega)=\sum_P \mathcal{A}_P(\omega).
 \label{eq:path_sum}
\end{equation}
Equation~\eqref{eq:path_sum} will be central in the channel-resolved analysis introduced in Sec.~\ref{sec:channels}, because it allows the total THG signal to be decomposed into pathway weights and interference terms.

\section{Numerical procedure}
\label{sec:numerics}

The bound-state problem defined by Eq.~\eqref{eq:chi_eq} was solved numerically through a finite-difference discretization on a radial mesh together with a self-consistent iteration in the energy argument of $V_{\mathrm{eff}}(r;E)$. Because the effective potential contains inverse powers of $r$, the radial grid is initiated at a small positive cutoff $r_{\min}>0$ and extended up to a sufficiently large $r_{\max}$, where the bound-state wave functions are already negligible. Dirichlet boundary conditions are imposed at the two endpoints, and the radial domain, mesh spacing, and cutoff position are increased until the low-energy eigenvalues and the dominant dipole matrix elements remain stable within the numerical precision of the plots. At each iteration step, a trial energy is used to build the effective potential, the discretized eigenvalue problem is solved, and the trial value is updated until the input and output energies converge.

The plotted effective potential denotes the real part of $V_{\mathrm{eff}}(r;E)$ evaluated at the converged reference energy of the state under consideration. The imaginary terms generated by the mixed radial--azimuthal metric component are retained in the differential operator and in the eigenvalue calculation; they do not lead to complex eigenenergies because the original curved-space Hamiltonian is Hermitian with respect to the measure $\sqrt{g}\,dr\,d\varphi=r\,dr\,d\varphi$. Once the normalized radial eigenfunctions are obtained, dipole matrix elements are evaluated from Eq.~\eqref{eq:dipole_matrix}, and the nonlinear susceptibilities are computed from Eqs.~\eqref{eq:chi2} and \eqref{eq:chi3}. The THG sums retain the low-lying states satisfying the dipole rule $\Delta m=\pm1$; convergence of the spectra was checked by enlarging the retained-state basis until the positions and relative heights of the displayed resonances did not change appreciably.

Unless otherwise stated, the calculations adopt GaAs-inspired parameters for the effective mass and confinement strength, with $\mu=0.067m_e$ and a displaced parabolic potential centered at $r_0$. The reference value used in Figs.~\ref{fig:veff}--\ref{fig:Enm} is $r_0=800\,\mathrm{nm}$, with $\hbar\omega_0=0.449\,\mathrm{meV}$.

\section{Results and discussion}
\label{sec:results}

\subsection{Effective potential, radial probability density and energy spectrum}
\label{sec:effective_potential_results}

To understand how the topological defect alters the nonlinear optical response of the quantum ring, we must first assess its influence on the spatial confinement of charge carriers. The real part of the effective radial potential, in the absence of a magnetic field ($B=0$), as a function of distance $r$ for different values of the spiral displacement parameter $\beta$ is shown in Figure~\ref{fig:veff}. In the absence of any defect ($\beta=0$, black curve), the confinement reproduces the typical profile of a displaced harmonic oscillator, which is proportional to $(r-r_0)^2$ and has the mean radius of the ring centered at $r_0$ [Eq.~\eqref{eq:confinement}]. As $\beta$ increases, the mixed radial-angular terms, caused by torsion in Eqs.~\eqref{eq:Bcoef} and \eqref{eq:Qcoef}, deform the confinement profile. This geometric deformation maintains the principal minimum at $r=r_0$, while causing a second, inner well to appear in the region between $r=0$ and $r_0$. It is important to note that, as $\beta$ increases, the local minimum of this secondary well becomes deeper, coming ever closer to the energy value of the principal minimum.
\begin{figure*}[tbhp]
    \centering
    \includegraphics[width=0.8\linewidth]{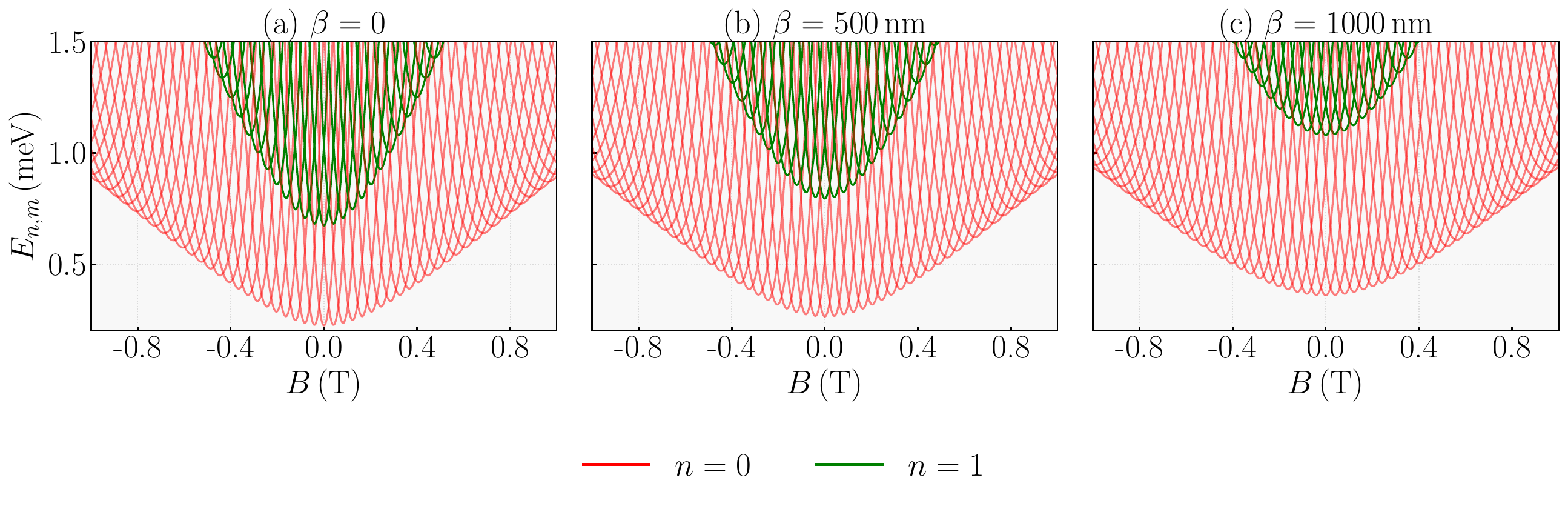}
    \caption{\footnotesize (Color online) Magnetic field dependence of the electronic energy levels for the first two radial subbands ($n=0$ and $n=1$) of the quantum ring. Within each colored set, the multiple thin curves correspond to the angular-momentum states $m$ retained in the calculation. The panels show selected spiral dislocation strengths: (a) $\beta = 0$, (b) $\beta = 500\,\mathrm{nm}$, and (c) $\beta = 1000\,\mathrm{nm}$.}
    \label{fig:Enm}
\end{figure*}

The changes induced by $\beta$ in the effective potential $V_{\mathrm{eff}}(r)$ with the emergence of the second internal well directly alter the profiles of the electron's wave functions. Fig.~\ref{fig:prob_density} shows the radial probability density distributions for the low-energy states. As the topological defect parameter $\beta$ increases, the peak amplitudes of the probability densities decrease. This decrease in amplitude occurs as a result of the creation of the inner well, which causes the wave function to spread out to adapt to the new, distorted confinement configuration. It is important to note that, although the point of maximum localization remains aligned with the average radius of the ring, the decrease in amplitude signals a new spatial distribution of the electron, reflecting a more extensive radial delocalization of the confined states.

The changes of the energy subbands are due to the reduction of the amplitudes of the probability densities induced by $\beta$. In Fig. \ref{fig:Enm} we present the dependence of the energy eigenvalues on the magnetic field and the evolution of the respective subbands as the parameter $\beta$ increases. It is seen that the minima of the two first subbands ($n=0$ and $n=1$) are shifted to higher energy levels as $\beta$ increases.

It is worth noting a striking feature regarding the scaling of the topological defect $\beta$ in our findings. In Figs.~\ref{fig:veff}, \ref{fig:prob_density}, and \ref{fig:Enm}, exaggerated values of $\beta$ (on the order of the ring radius, $r_0$) were required to produce visually discernible modifications in the effective potential profile, wave functions, and energy subbands. This occurs because the defect is located at the origin ($r=0$) and its influence decays rapidly with distance via the inverse radial terms (proportional to $1/r^2$ and $1/r^3$) in the effective potential. However, for the nonlinear optical properties discussed below, we restrict $\beta$ to a much smaller range ($0 - 10\,\mathrm{nm}$). Despite the structural deformations being visually imperceptible at this scale, the THG spectra exhibit drastic changes. This hyper-sensitivity arises from the highly resonant denominators in the susceptibility $\chi^{(3)}$ (Eq.~\eqref{eq:Dabc}): since the dephasing broadening is small ($\Gamma = 0.10\,\mathrm{meV}$), even minimal variations in transition energies---amplified by the quadruple product of the dipole matrix elements---are sufficient to completely shift and rearrange the THG peaks. Therefore, our model suggests that the multiresonant nonlinear optical response acts as an extremely sensitive probe for identifying topological defects.

\subsection{Dipole-allowed transitions, and THG spectra}
\label{sec:thg_spectra_results}

To address nonlinear spectra, it is useful to summarize the structure of the low-energy transitions that make up the denominators of the THG in Eq.~\eqref{eq:Dabc}. Table~\ref{tab:transition_energies} shows the set of typical transition energies for the dipole-allowed channels. It illustrates how the state-dependent energy shifts evolve as $\beta$ grows, and clarifies which energy gaps underlie the resonances marked on the THG plots. In this context, these energy disparities are not merely quantitative measures: they define where one-, two-, and three-photon conditions can arise and thus determine the entire spectrum of $\chi^{(3)}$.

Furthermore, Table~\ref{tab:transition_energies} reveals another physical characteristic for the case of a finite magnetic field ($B=1\,\mathrm{T}$). For larger values of distortion, such as $\beta=6.0$ and $9.0\,\mathrm{nm}$, the values of $\Delta E$ for the $(0,0)\rightarrow(0,1)$ transition become slightly negative. This signals a magnetic-field-induced level inversion in which the state with angular momentum $m=1$ becomes energetically lower than the $m=0$ state. In the spectra below, $\ket{0,0}$ is used as a reference initial state in order to track the same transition pathways continuously as $\beta$ and $B$ are varied. When this inversion occurs, the corresponding curves should therefore be interpreted as the nonlinear response of an optically prepared reference state rather than as a strictly thermal-equilibrium ground-state response. A thermal-equilibrium calculation in that parameter region would require resetting $\ket{g}$ to the true lowest-energy state.

\begin{table}[htpb]
\centering
\caption{Representative transition energies $\Delta E$ (in meV) for the dipole-allowed channels that compose the THG resonance pathways. The values are presented for both zero ($B=0$) and finite ($B=1\,\mathrm{T}$) magnetic fields, explicitly showing the state-dependent shifts induced by the spiral dislocation.}
\label{tab:transition_energies}
\begin{tabular}{cccc}
\toprule
Transition & $\beta$ & $\Delta E$ ($B=0$) & $\Delta E$ ($B=1\,\mathrm{T}$) \\
$(n,m) \rightarrow (n',m')$ & (nm) & (meV) & (meV) \\
\midrule
$(0,0) \rightarrow (0,1)$ & \multirow{3}{*}{0.0} &  1.0703 &  0.2464 \\
$(0,1) \rightarrow (1,0)$ &                      &  5.7296 &  6.5310 \\
$(1,0) \rightarrow (1,1)$ &                      &  2.5579 &  1.7093 \\
\midrule
$(0,0) \rightarrow (0,1)$ & \multirow{3}{*}{3.0} &  0.9327 &  0.1082 \\
$(0,1) \rightarrow (1,0)$ &                      &  6.6863 &  7.4906 \\
$(1,0) \rightarrow (1,1)$ &                      &  1.8544 &  1.0189 \\
\midrule
$(0,0) \rightarrow (0,1)$ & \multirow{3}{*}{6.0} &  0.8004 & -0.0031 \\
$(0,1) \rightarrow (1,0)$ &                      &  7.8352 &  8.6217 \\
$(1,0) \rightarrow (1,1)$ &                      &  1.3424 &  0.5503 \\
\midrule
$(0,0) \rightarrow (0,1)$ & \multirow{3}{*}{9.0} &  0.6937 & -0.0742 \\
$(0,1) \rightarrow (1,0)$ &                      &  9.0404 &  9.7945 \\
$(1,0) \rightarrow (1,1)$ &                      &  1.0285 &  0.2889 \\
\bottomrule
\end{tabular}
\end{table}

\begin{figure}[tbhp]
\centering
\includegraphics[width=\linewidth]{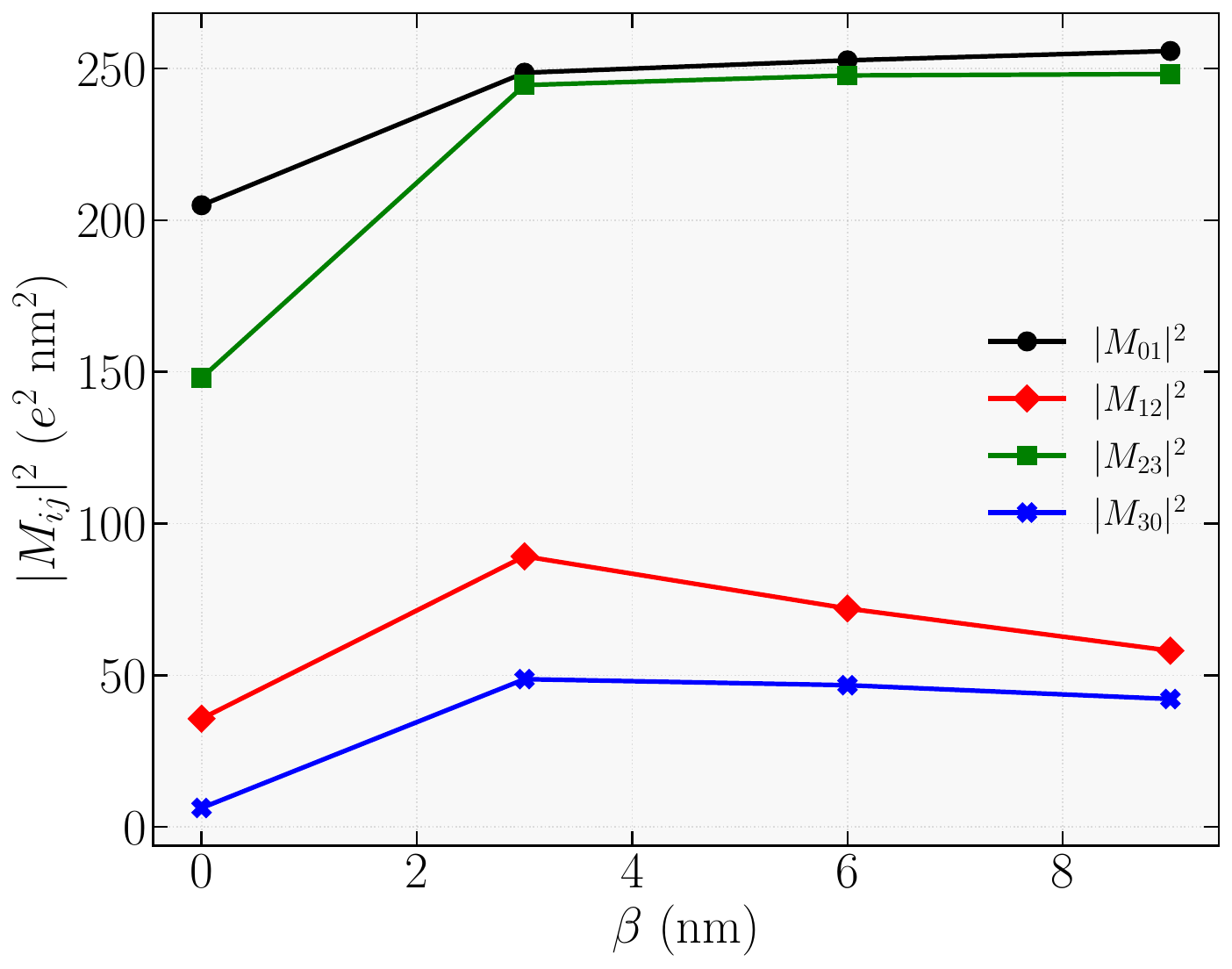}
\caption{\footnotesize (Color online) Squared dipole matrix elements $|M_{ij}|^2$ for the representative dipole-allowed THG path $g\rightarrow a\rightarrow b\rightarrow c\rightarrow g$, with $g=(0,0)$, $a=(0,1)$, $b=(1,0)$, and $c=(1,1)$. The labels $M_{01}$, $M_{12}$, $M_{23}$, and $M_{30}$ refer to these consecutive links.}
\label{fig:matrix_elements}
\end{figure}

The next step is to examine how the defect alters the dipole couplings that enter Eq.~\eqref{eq:chi3}. Figure~\ref{fig:matrix_elements} collects the squared dipole moments for a representative allowed THG chain. In that plot, $M_{01}$, $M_{12}$, $M_{23}$, and $M_{30}$ denote the four consecutive dipole factors entering the path $g\rightarrow a\rightarrow b\rightarrow c\rightarrow g$, with $g=(0,0)$, $a=(0,1)$, $b=(1,0)$, and $c=(1,1)$. The important message is that the dislocation does not act as a simple global attenuation factor. Instead, $\beta$ changes the radial overlap in a transition-dependent manner, so some channels are enhanced, others are weakened, and some may even display non-monotonic behavior. This state selectivity is one of the reasons why the THG signal is so sensitive to geometry: the susceptibility contains a product of four dipole elements and therefore responds strongly to comparatively modest changes in individual overlaps.

\begin{figure}[tbhp]
\centering
\includegraphics[width=\linewidth]{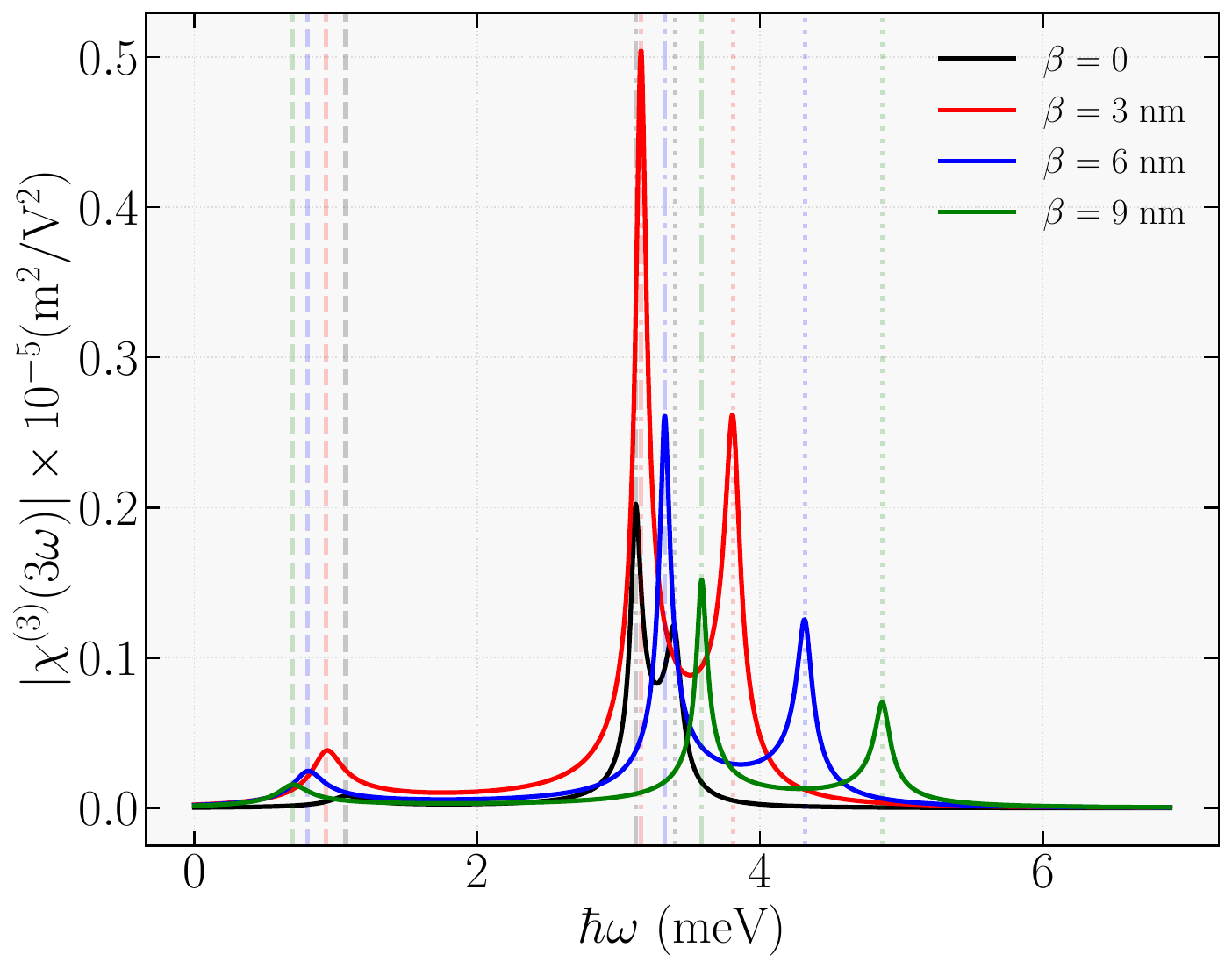}
\caption{\footnotesize (Color online) Third-harmonic generation susceptibility $|\chi^{(3)}(3\omega)|$ versus photon energy $\hbar\omega$ for different values of $\beta$ in zero magnetic field ($B = 0$). We set the dephasing broadening to be $\Gamma = 0.10\,\mathrm{meV}$.}
\label{fig:thg_B0}
\end{figure}

\begin{figure}[tbhp]
\centering
\includegraphics[width=\linewidth]{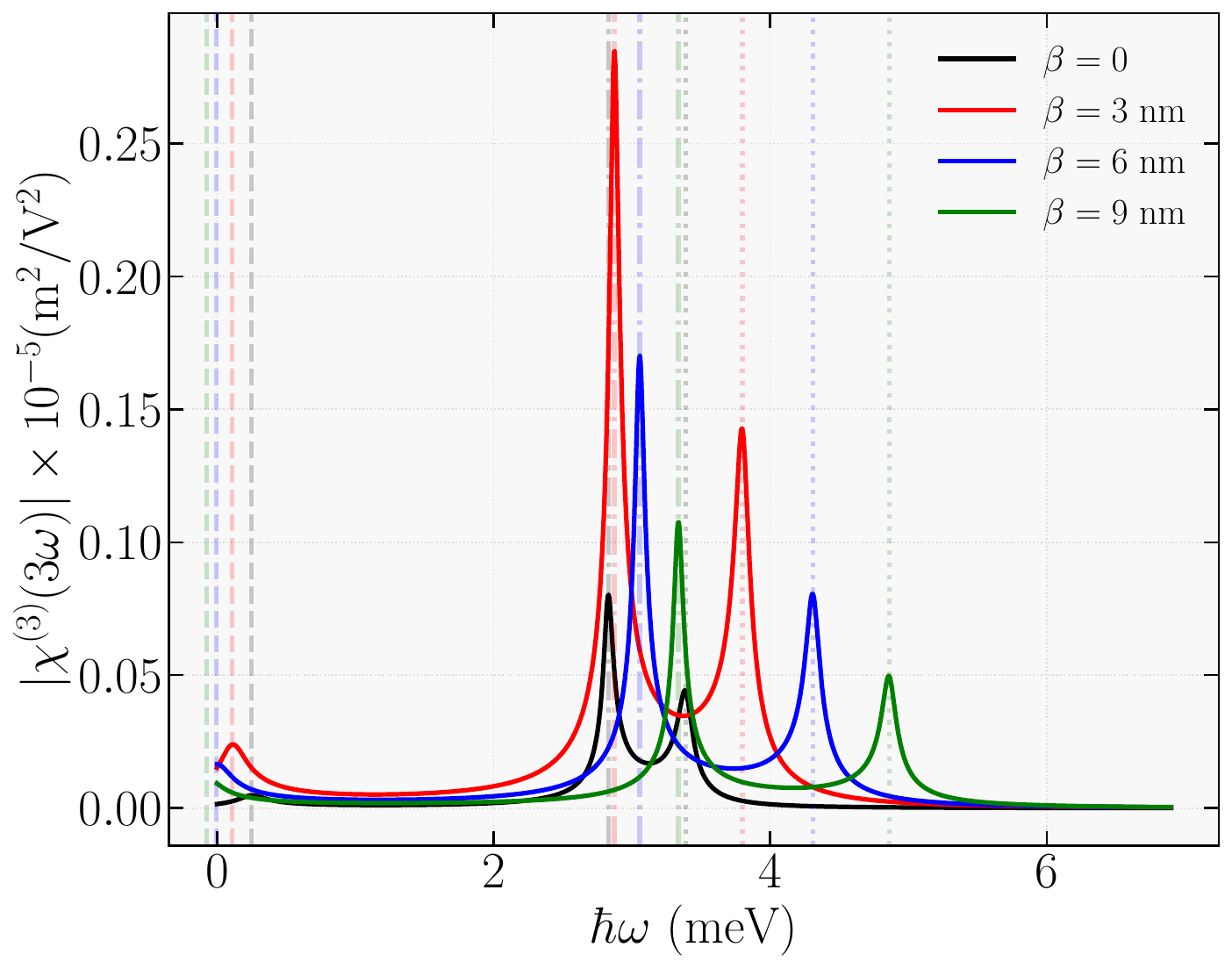}
\caption{\footnotesize (Color online) Third-harmonic generation susceptibility $|\chi^{(3)}(3\omega)|$ versus photon energy $\hbar\omega$ for different values of $\beta$ at finite magnetic field $B = 1\,\mathrm{T}$.}
\label{fig:thg_B1}
\end{figure}

The THG spectra are shown in Figures~\ref{fig:thg_B0} and \ref{fig:thg_B1} for magnetic fields of $B=0$ and $B=1\mathrm{T}$, respectively, for different values of $\beta$. The spectra are therefore inherently multiresonant, since the denominator in Eq.~\eqref{eq:Dabc} includes one-, two-, and three-photon resonances. Changing $\beta$ reorders the positions of the dominant structures and alters their intensities. In practical terms, the shift alters both the resonance energies across the spectrum and the peak intensities, through the quadruple product of the elements of the dipole matrix, $M_{g a} M_{a b} M_{b c} M_{c g}$. The resulting tendency reveals a complex state-dependent reorganization of the energy spectrum: while the lowest-energy resonance undergoes a red-shift as $\beta$ increases, the higher-energy multiphoton resonances are pushed toward higher values (blue-shift). The overall spectral intensity is governed by the competition between reduced detuning and altered wave-function overlap.

\begin{figure}[tbhp]
\centering
\includegraphics[width=\linewidth]{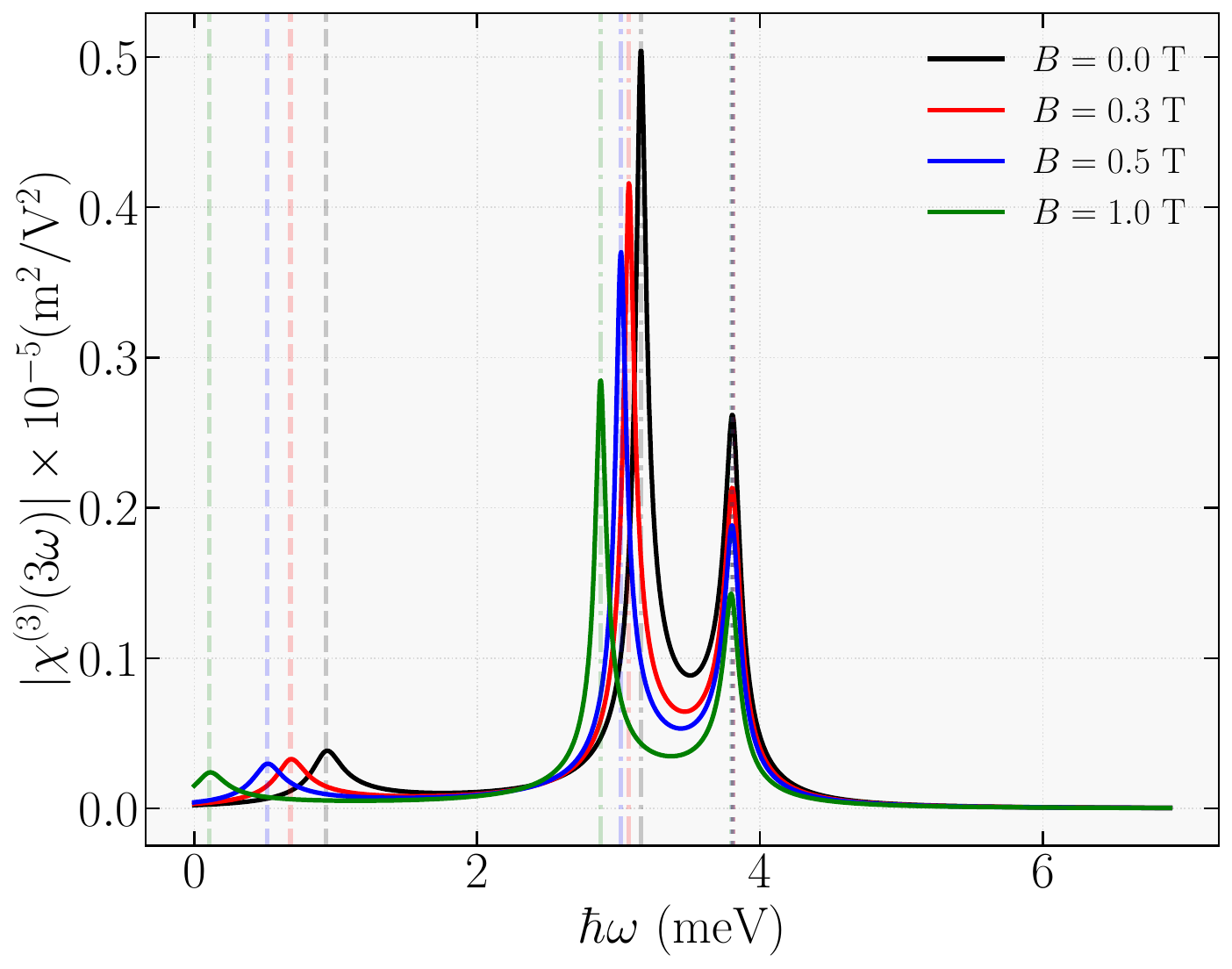}
\caption{\footnotesize (Color online) Third-harmonic generation susceptibility $|\chi^{(3)}(3\omega)|$ versus the photon energy $\hbar\omega$ for different values of the magnetic field $B$ fixing the spiral dislocation strength in $\beta=3\,\mathrm{nm}$.}
\label{fig:thg_beta_fixed}
\end{figure}

At fixed dislocation strength, the magnetic field provides an additional control knob. The spectra in Fig.~\ref{fig:thg_beta_fixed} indicate that $B$ changes the separation between resonant structures and can modify their relative intensities in a non-monotonic way. This behavior is expected because the field alters the orbital energies and hence the detunings in Eq.~\eqref{eq:Dabc}, while also changing the underlying wave-function overlaps. In qualitative terms, rather than acting in opposition to the geometric defect, the magnetic field intensifies the state-dependent spectral reorganization: increasing $B$ further pushes the lowest-energy resonance toward lower values (red-shift) while driving the higher-energy structures to even higher values (blue-shift). The combined effect of these interactions dramatically modifies the effective orbital confinement. This interplay is an indicator that $\beta$ and $B$ play highly complementary roles in tailoring the nonlinear optical response.

\subsection{Extension I: dephasing-controlled spectral resolution}
\label{sec:dephasing}

The existing THG figures already suggest that the apparent number of peaks depends not only on the transition energies but also on the linewidth parameter. This motivates an explicit dephasing analysis. In the density-matrix equation, the microscopic dephasing rate is written as
\begin{equation}
 \widetilde{\Gamma}_{ij}=\frac{1}{T_{2,ij}},
 \label{eq:dephasing_rate}
\end{equation}
where $T_{2,ij}$ is the coherence time of the transition. In the optical spectra, however, it is more convenient to use the corresponding energy broadening
\begin{equation}
 \Gamma_{ij}=\hbar\widetilde{\Gamma}_{ij}=\frac{\hbar}{T_{2,ij}}.
 \label{eq:energy_broadening}
\end{equation}
Throughout the numerical results, $\Gamma$ denotes this dephasing broadening in energy units, expressed in meV. When one assumes, for simplicity, a common coherence time $T_2$ for the dominant optical transitions, one has $\Gamma_{ag}=\Gamma_{bg}=\Gamma_{cg}\equiv\Gamma$. For the representative value $T_2=15\,\mathrm{ps}$, one obtains $\Gamma\approx 0.044\,\mathrm{meV}$, which is small enough to resolve nearby resonances in typical THG spectra.

A useful extension of the numerical analysis is therefore to compute $|\chi^{(3)}(3\omega)|$ for several values of $\Gamma$ (or, equivalently, $T_2$) while keeping $\beta$ and $B$ fixed. This produces two direct diagnostics. First, the spectral line shape evolves from well-resolved multiplet structures at weak dephasing to merged and broader envelopes at stronger dephasing. Second, the peak amplitude, full width at half maximum (FWHM), and spectral contrast can be tracked quantitatively as functions of $\Gamma$. Because the THG denominator contains three resonant factors, dephasing controls not only the linewidths but also the balance between overlapping one-, two-, and three-photon pathways.

\begin{figure}[tbhp]
\centering
\includegraphics[width=\linewidth]{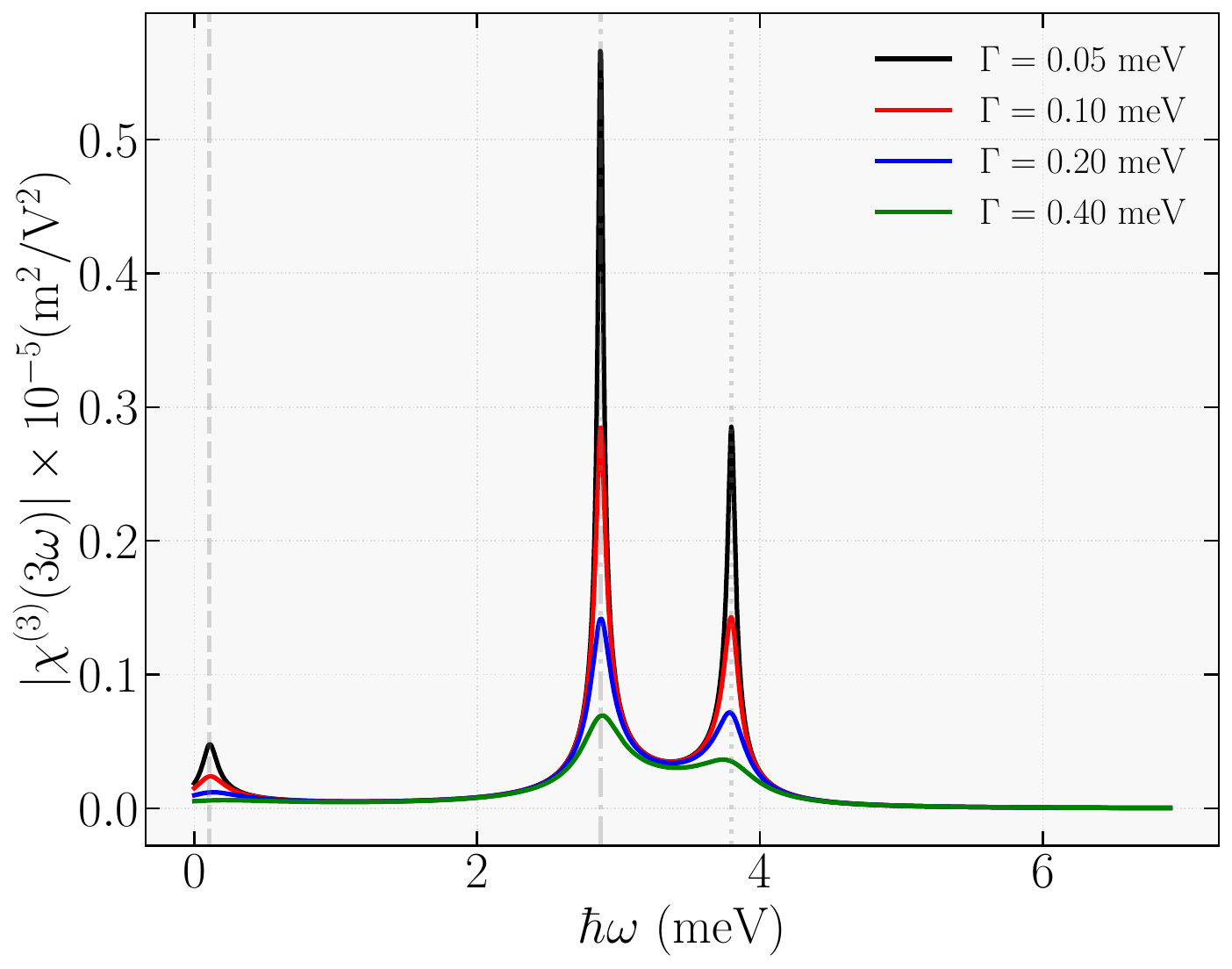}
\caption{\footnotesize (Color online) Third-harmonic generation susceptibility $|\chi^{(3)}(3\omega)|$ as a function of the photon energy $\hbar\omega$, for representative values of the dephasing broadening $\Gamma$, at fixed $\beta = 3.0\,\mathrm{nm}$ and $B = 1.0\,\mathrm{T}$.}
\label{fig:dephasing_scan}
\end{figure}

Figure~\ref{fig:dephasing_scan} presents the THG spectrum computed for $\beta = 3.0\,\mathrm{nm}$ and $B = 1.0\,\mathrm{T}$ with representative values of the dephasing broadening, ranging from $\Gamma = 0.05\,\mathrm{meV}$ to $\Gamma = 0.40\,\mathrm{meV}$. At the lowest dephasing broadening considered ($\Gamma = 0.05\,\mathrm{meV}$), the spectrum displays sharply resolved resonances with substantially larger amplitudes, whose positions correspond to one-, two-, and three-photon conditions consistent with the transition hierarchy summarized in Table~\ref{tab:transition_energies}. As $\Gamma$ increases to intermediate values (such as $0.10$ and $0.20\,\mathrm{meV}$), the individual peaks broaden, their intensities drop severely, and neighboring resonances begin to overlap. At the largest value in the range ($\Gamma = 0.40\,\mathrm{meV}$), the resonances have merged completely, and the spectral profile is dominated by a smooth, broad maximum with reduced contrast. The peak amplitude drops by roughly one order of magnitude across the range of $\Gamma$ values explored, confirming that the THG signal is extremely sensitive to the coherence properties of the system. This behavior is physically expected: the denominator in Eq.~\eqref{eq:Dabc} contains three resonant factors, so the peak height scales approximately as $\Gamma^{-3}$ in the limit of well-separated resonances, making THG considerably more dephasing-sensitive than lower-order processes.

\subsection{Extension II: two-parameter spectral evolution}
\label{sec:maps}

\begin{figure*}[tbhp]
\centering
\includegraphics[width=0.45\linewidth]{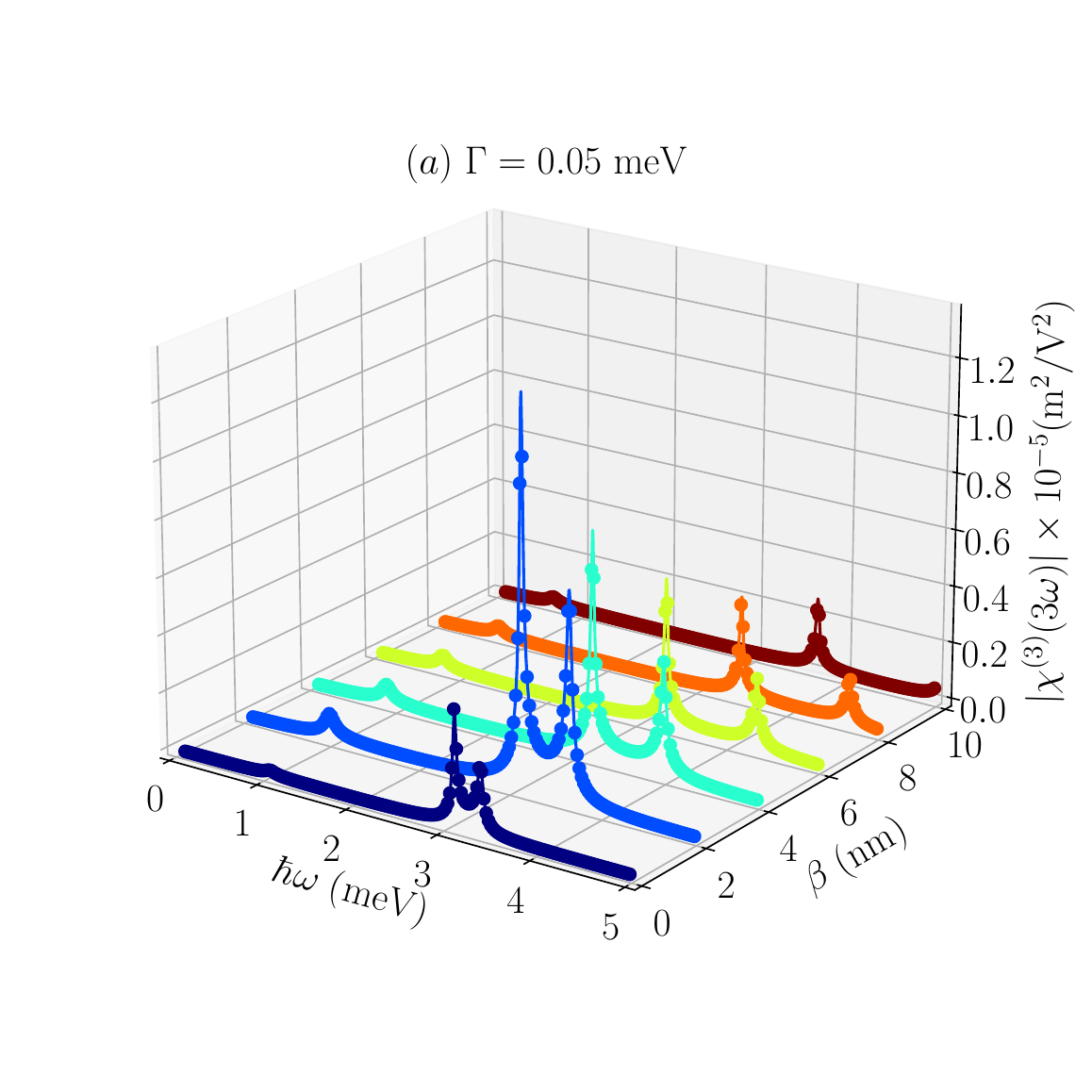}
\includegraphics[width=0.45\linewidth]{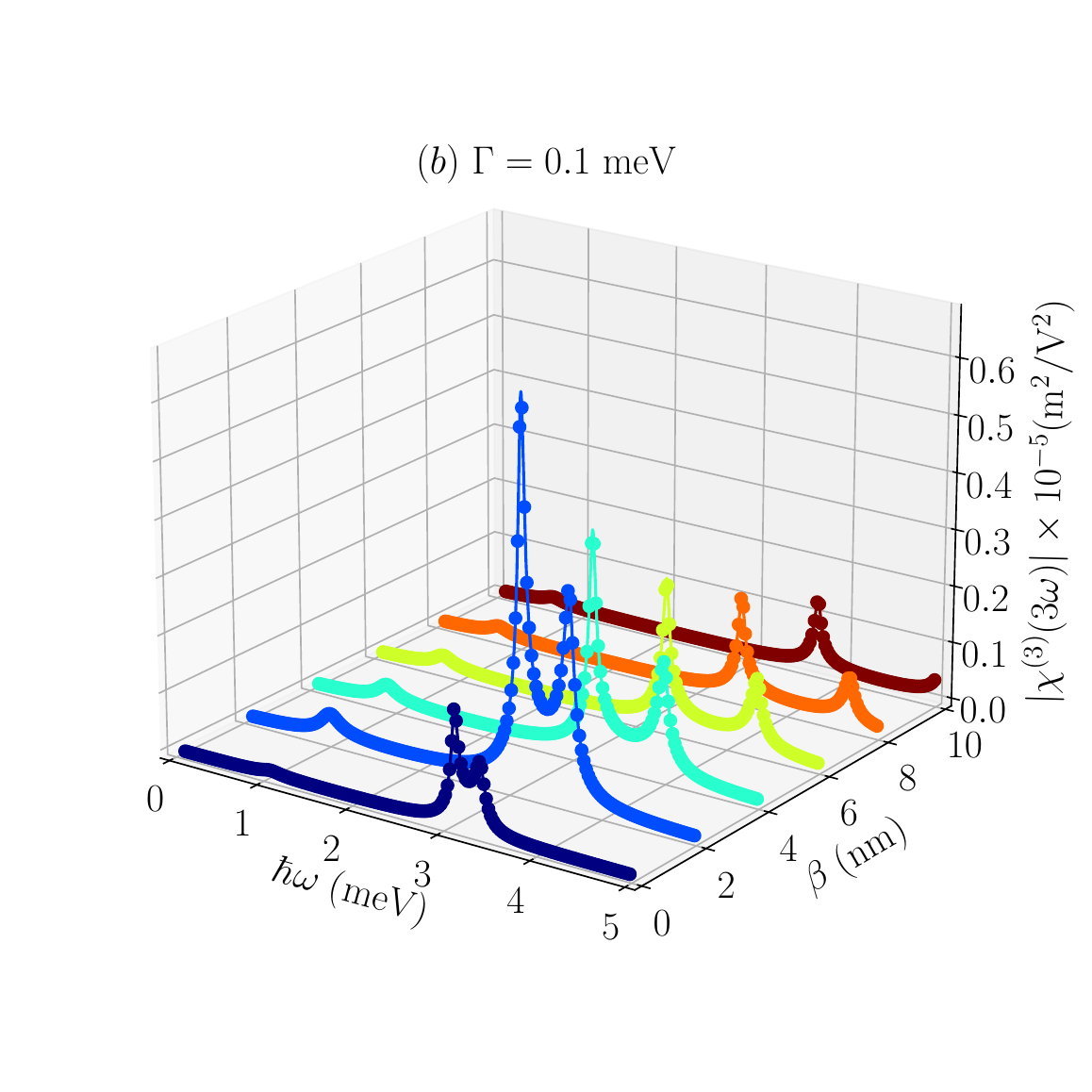}\\
\includegraphics[width=0.45\linewidth]{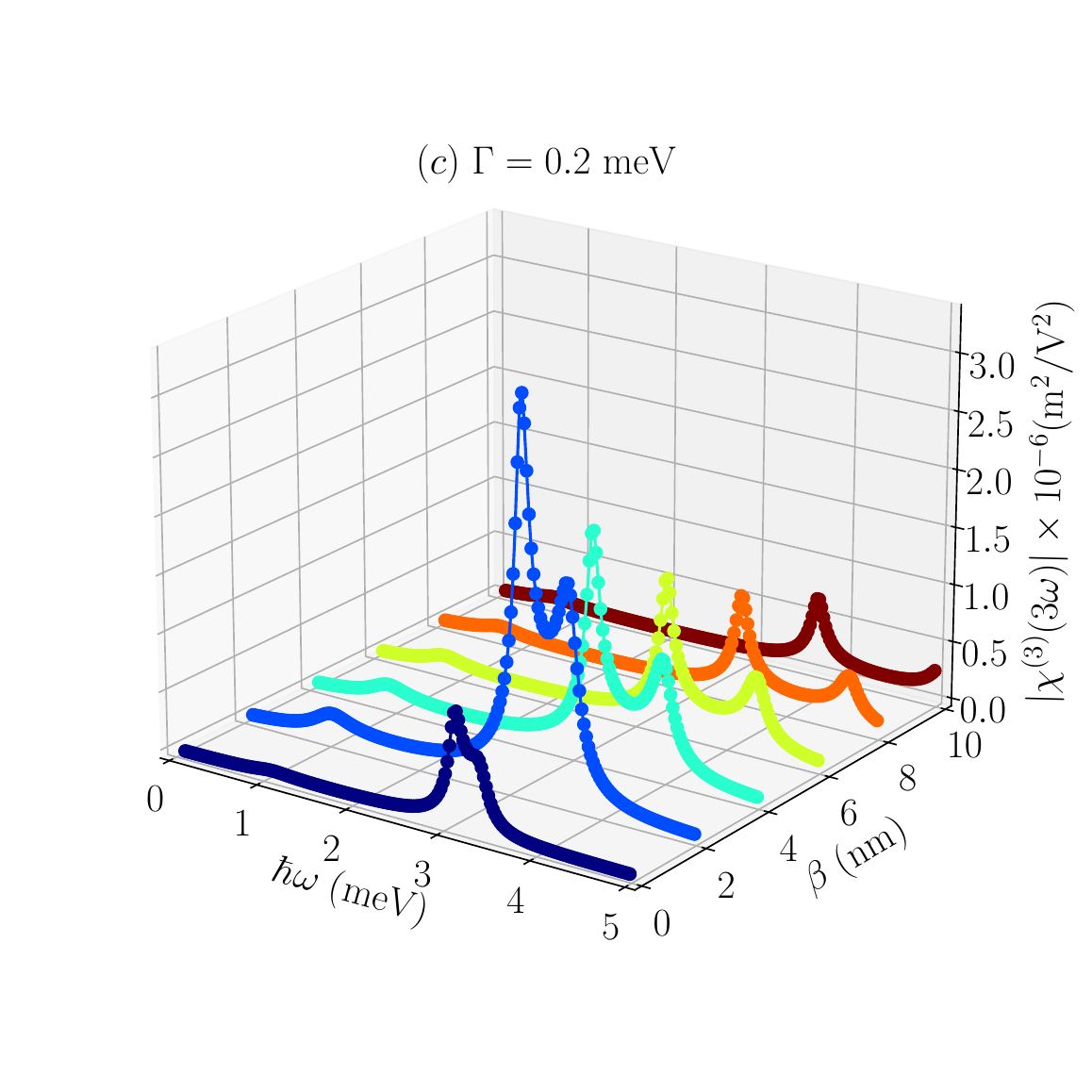}
\includegraphics[width=0.45\linewidth]{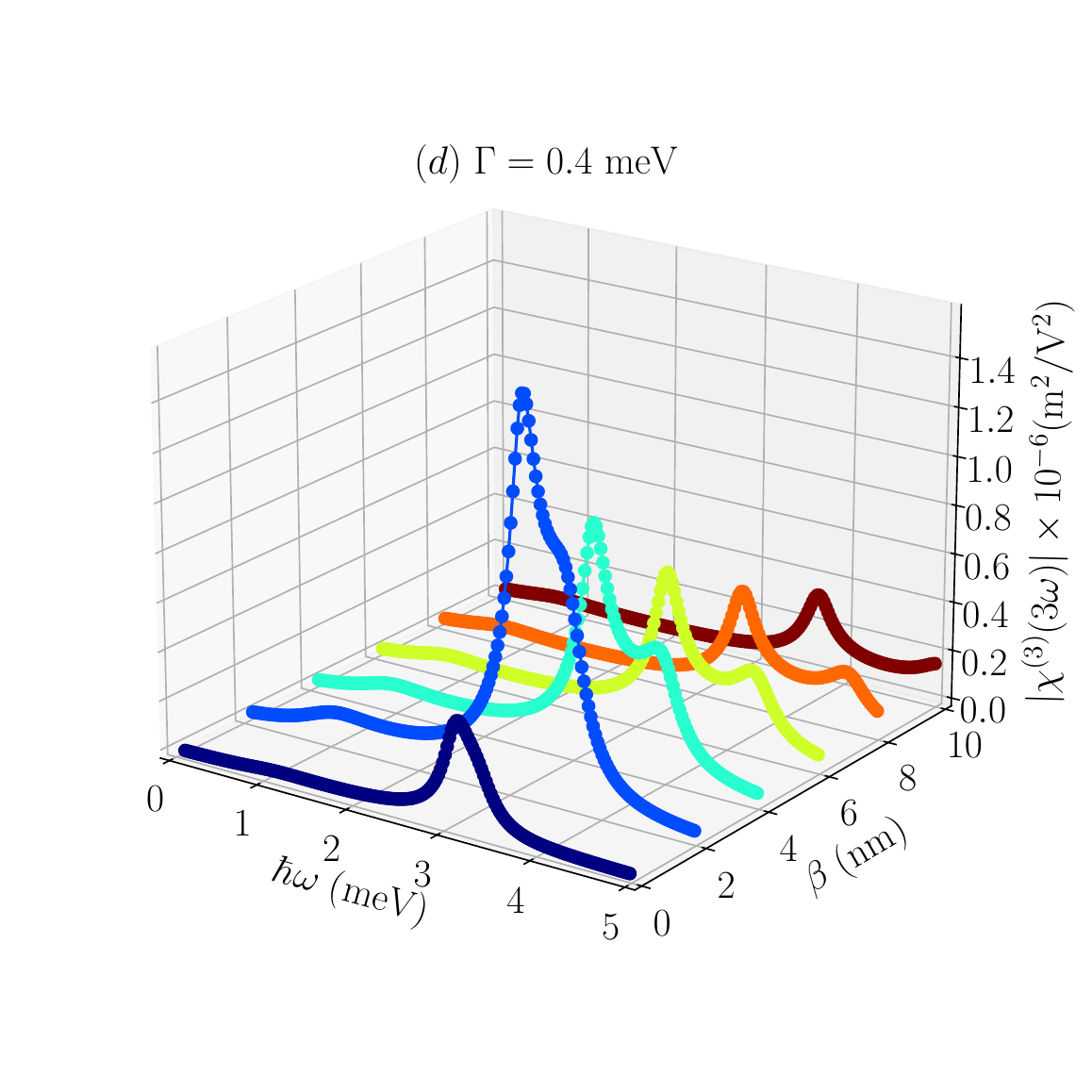}
\caption{\footnotesize (Color online) Third-harmonic generation susceptibility versus the photon energy $\hbar\omega$ and the spiral dislocation parameter $\beta \in \{0, 2, 4, 6, 8, 10\}\,\mathrm{nm}$. The panels show the results for fixed dephasing broadenings: (a) $\Gamma=0.05\,\mathrm{meV}$, (b) $\Gamma=0.10\,\mathrm{meV}$, (c) $\Gamma=0.20\,\mathrm{meV}$ and (d) $\Gamma=0.40\,\mathrm{meV}$.}
\label{fig:thg_waterfallG}
\end{figure*}

To get an overview of the changes in the THG spectrum caused by the topological defect for different values of $\Gamma$, Fig.~\ref{fig:thg_waterfallG} presents a 3D waterfall plot of the THG susceptibility (z-axis) as a function of photon energy $\hbar\omega$ (x-axis) and the spiral displacement parameter $\beta \in \{0, 2, 4, 6, 8, 10\}\,\mathrm{nm}$ (y-axis). Panels (a) through (d) show various dephasing broadenings, ranging from $\Gamma=0.05$ to $\Gamma=0.40\,\mathrm{meV}$. This waterfall plot captures the interplay between the constant shift in resonance positions induced by $\beta$ and the amplitude attenuation defined by $\Gamma$.

\begin{figure*}[tbhp]
\centering
\includegraphics[width=0.45\linewidth]{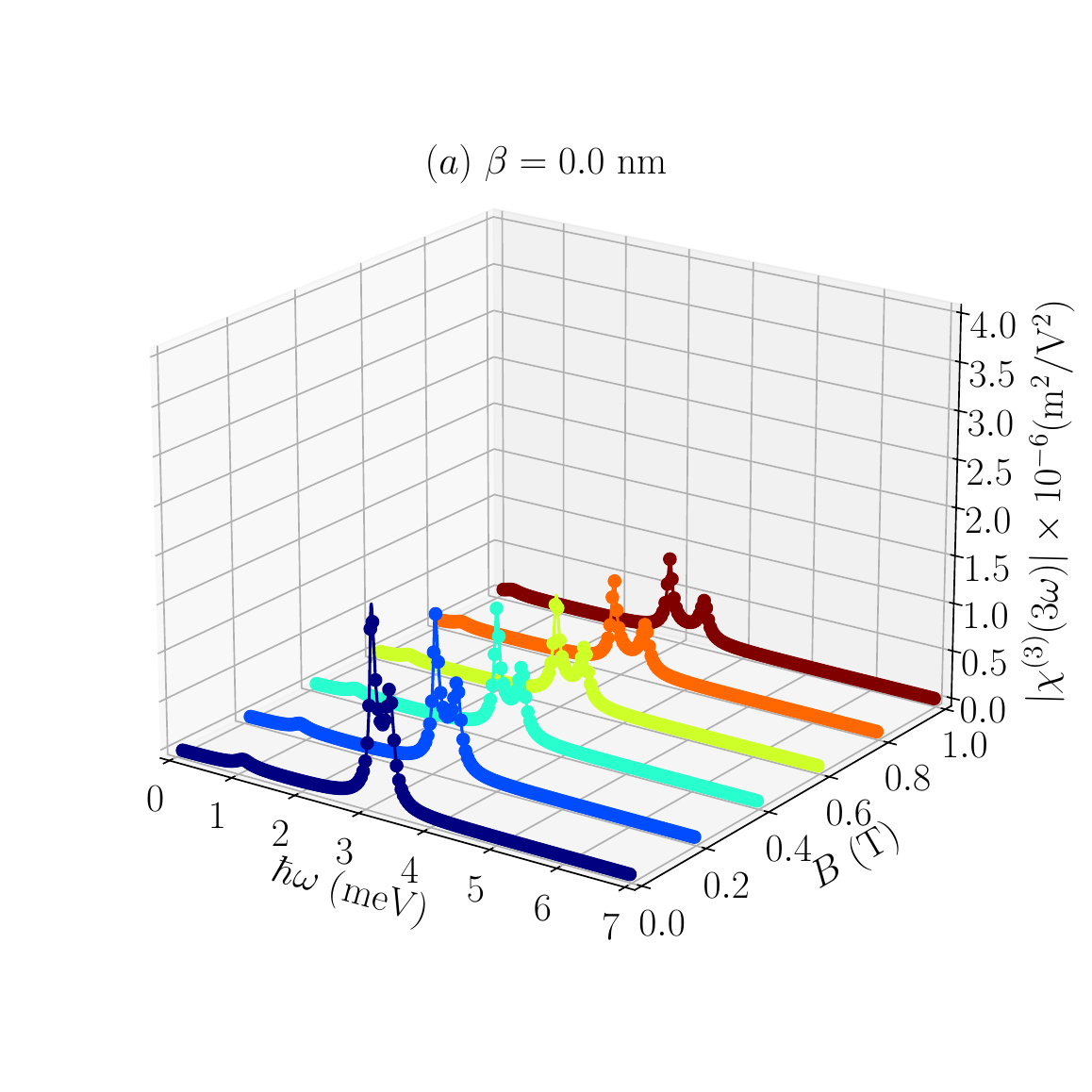}
\includegraphics[width=0.45\linewidth]{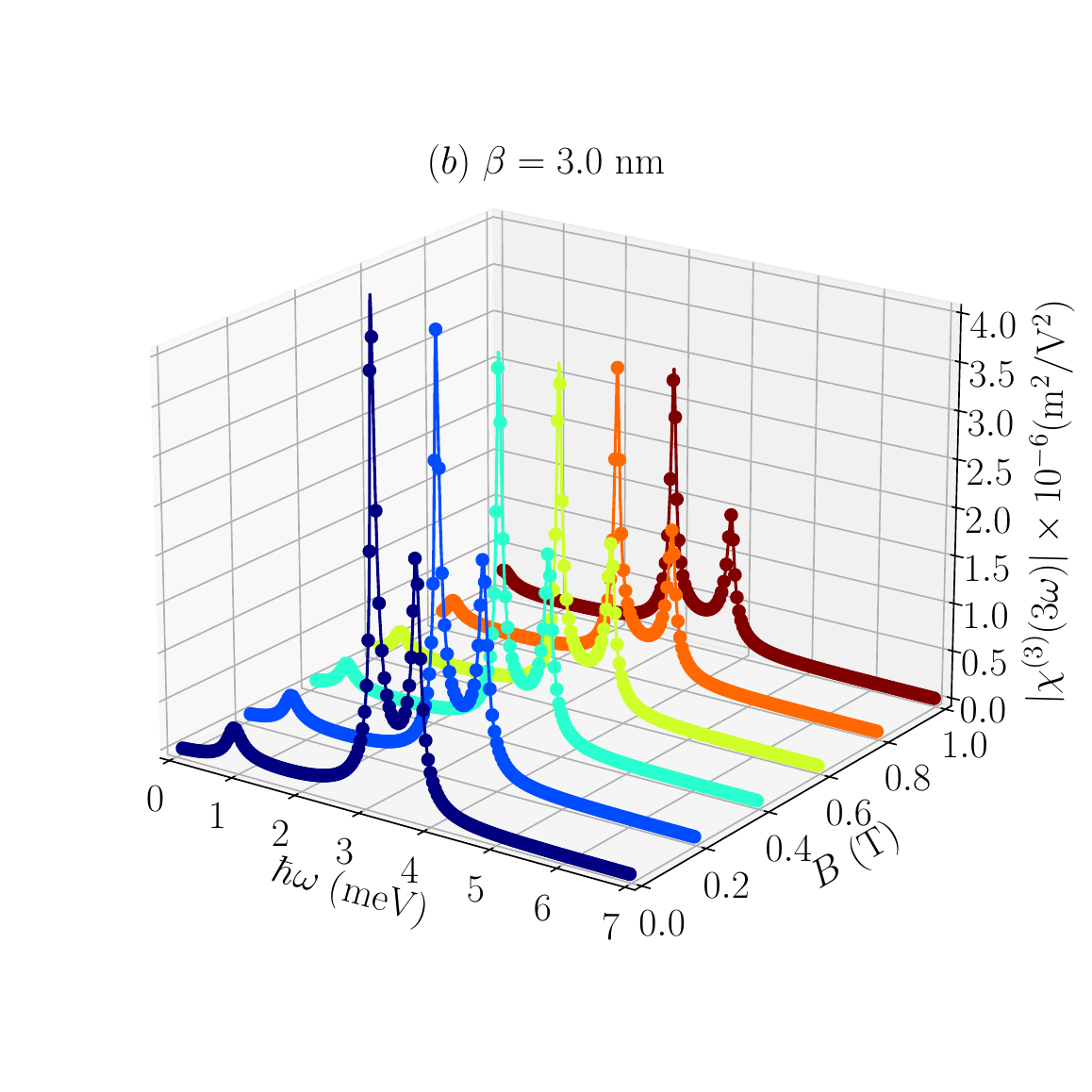}\\
\includegraphics[width=0.45\linewidth]{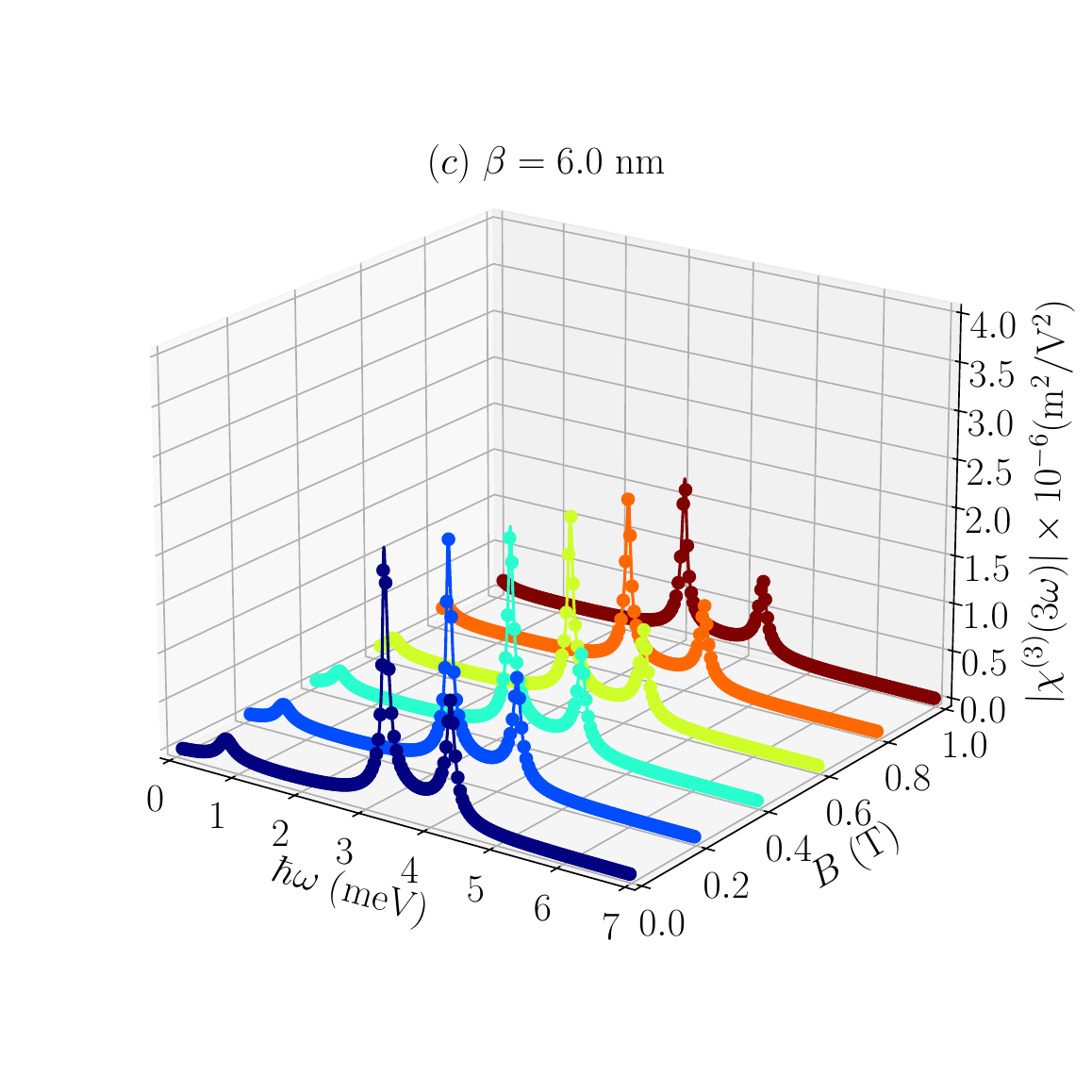}
\includegraphics[width=0.45\linewidth]{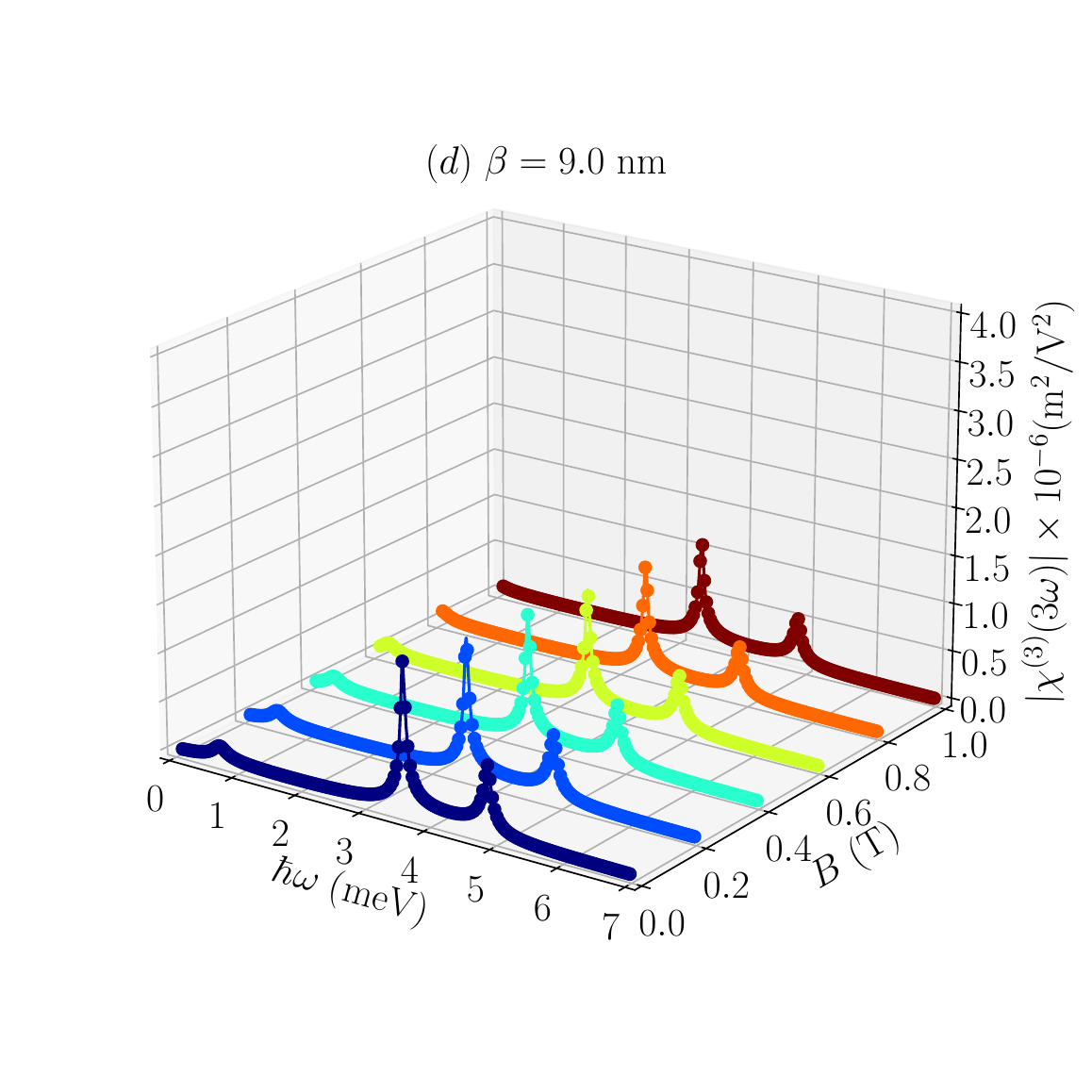}
\caption{\footnotesize (Color online) Magnetic field dependence of the third-harmonic generation susceptibility as a function of the photon energy $\hbar\omega$ for $B \in \{0, 0.2, 0.4, 0.6, 0.8, 1.0\}\,\mathrm{T}$. The panels show the effects for fixed spiral dislocation parameters: (a) $\beta=0\,\mathrm{nm}$, (b) $\beta=3\,\mathrm{nm}$, (c) $\beta=6\,\mathrm{nm}$ and (d) $\beta=9\,\mathrm{nm}$.}
\label{fig:thg_waterfallB}
\end{figure*}

Having established the sensitivity of the THG signal to dephasing, we turn our attention to the combined interplay between geometry and magnetic confinement. Fig.~\ref{fig:thg_waterfallB} presents the THG susceptibility (z-axis) as a function of the photon energy $\hbar\omega$ (x-axis) and the magnetic field $B \in \{0, 0.2, 0.4, 0.6, 0.8, 1.0\}\,\mathrm{T}$ (y-axis). The panels correspond to fixed values of the spiral dislocation parameter: (a) $\beta=0\,\mathrm{nm}$, (b) $\beta=3\,\mathrm{nm}$, (c) $\beta=6\,\mathrm{nm}$, and (d) $\beta=9\,\mathrm{nm}$. It is observed that an increase in $B$ drastically suppresses the amplitudes of the THG resonance peaks. Furthermore, as $\beta$ increases, this attenuation of the amplitudes becomes even more pronounced, which is consistent with the results presented in Figs.~\ref{fig:thg_B0} and \ref{fig:thg_beta_fixed}.

\subsection{Extension III: channel-resolved THG decomposition}
\label{sec:channels}

The third extension is a pathway analysis based on Eqs.~\eqref{eq:pathA}--\eqref{eq:pathBneg}. Instead of plotting only the total modulus $|\chi^{(3)}|$, one can separate the coherent contributions associated with distinct angular-momentum chains. Let us denote by $\chi_A^{(3)}$, $\chi_B^{(3)}$, $\chi_{-A}^{(3)}$, and $\chi_{-B}^{(3)}$ the amplitudes associated with the four representative families in Eqs.~\eqref{eq:pathA}--\eqref{eq:pathBneg}. The total response is then
\begin{equation}
\begin{split}
 \chi^{(3)}={}&\chi_A^{(3)}+\chi_B^{(3)}
 +\chi_{-A}^{(3)}+\chi_{-B}^{(3)} +\chi_{\mathrm{rest}}^{(3)} .
\end{split}
 \label{eq:chi3_decomp}
\end{equation}
where $\chi_{\mathrm{rest}}^{(3)}$ includes additional allowed channels if more states are retained. The measurable intensity can be decomposed as
\begin{equation}
\begin{split}
 |\chi^{(3)}|^2={}&\sum_P |\chi_P^{(3)}|^2 +2\sum_{P<Q}\mathrm{Re}
 \left[\chi_P^{(3)}\chi_Q^{(3)*}\right] .
\end{split}
 \label{eq:interference_decomp}
\end{equation}
Equation~\eqref{eq:interference_decomp} makes transparent which part of the signal comes from individual pathways and which part comes from constructive or destructive interference.

This analysis is physically important because the spiral dislocation acts differently on intermediate states with different angular content. Even when two pathways share the same initial and final states, their radial overlaps and detunings need not respond equally to $\beta$ or $B$. A channel-resolved plot can therefore reveal whether the defect primarily changes the dominant path itself or instead changes the interference between competing paths. That information is not accessible from the total spectrum alone.

\begin{figure}[tbhp]
\centering
\includegraphics[width=\linewidth]{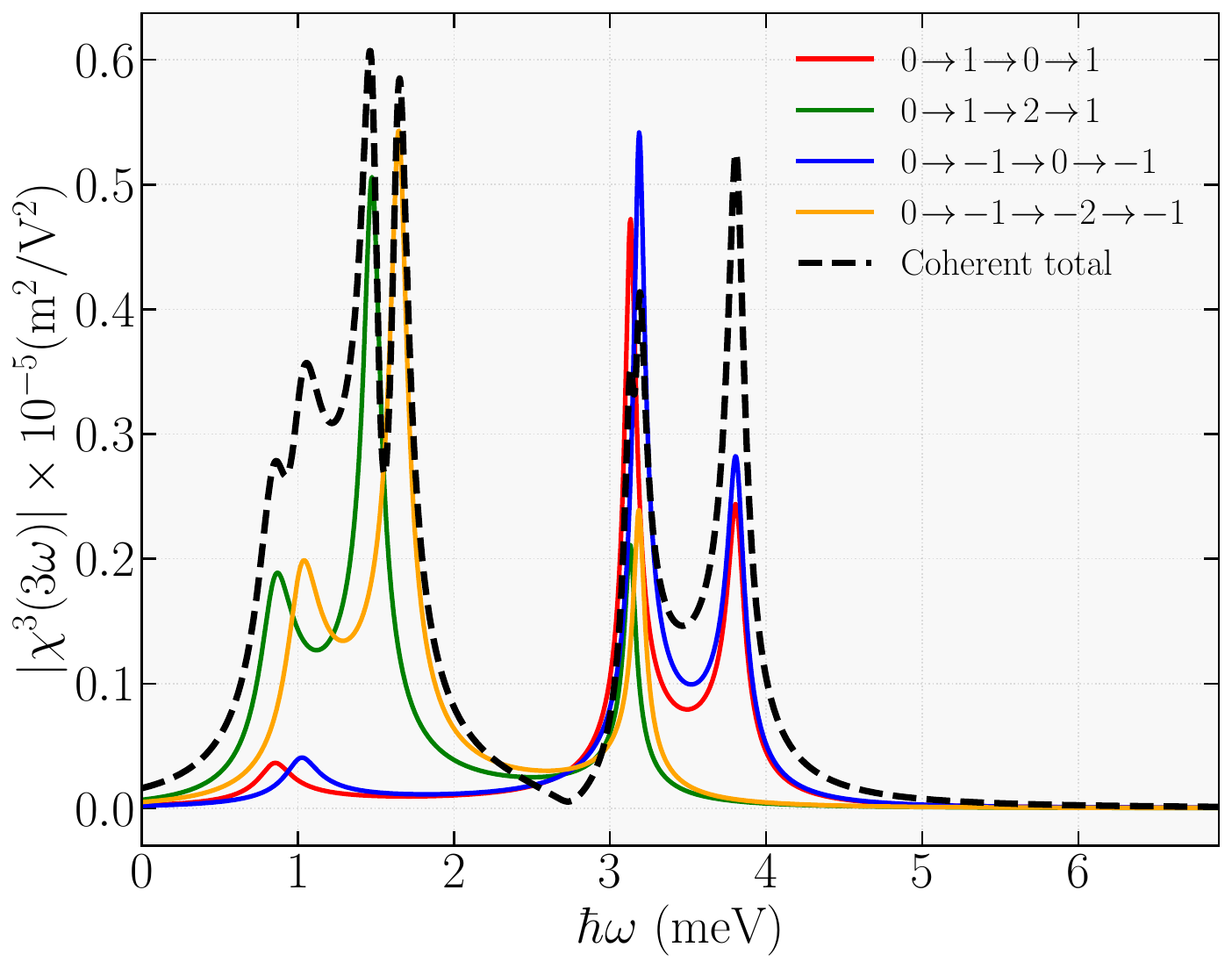}
\caption{\footnotesize (Color online) Channel-resolved decomposition of third-harmonic generation susceptibility at $\beta = 3.0\,\mathrm{nm}$, $B = 0.1\,\mathrm{T}$ and $\Gamma = 0.10\,\mathrm{meV}$. The solid coloured curves are the individual symmetry allowed transition chains and the dashed black curve is the coherent total.}
\label{fig:channel_decomp}
\end{figure}

Figure~\ref{fig:channel_decomp} presents the channel-resolved decomposition of the THG spectrum computed at $\beta=3.0\,\mathrm{nm}$, $B=0.1\,\mathrm{T}$, and $\Gamma=0.10\,\mathrm{meV}$. The individual contributions from the four symmetry-allowed transition chains are shown alongside the coherent total. Several features are immediately apparent. First, the two families that involve the intermediate state $\ket{n_2,0}$ (channels $0\!\to\!1\!\to\!0\!\to\!1$ and $0\!\to\!{-1}\!\to\!0\!\to\!{-1}$) produce the dominant spectral weight, with their resonance positions largely determined by the energy gaps $E_{(0,1)}-E_{(0,0)}$ and $E_{(1,0)}-E_{(0,0)}$. This behavior confirms the hierarchy of the transition energies summarized in Table~\ref{tab:transition_energies} above and verifies that these particular pathways are the dominant ones for the strong constructive interference observed at lower photon energies. Second, the channels that pass through the $|m|=2$ states ($0\!\to\!1\!\to\!2\!\to\!1$ and $0\!\to\!{-1}\!\to\!{-2}\!\to\!{-1}$) contribute at higher photon energies, consistent with the larger energy denominators associated with the second angular-momentum shell. Third, the coherent total (dashed black curve) does not coincide with the envelope of the individual contributions, indicating significant constructive and destructive interference between pathways in different spectral windows. In particular, near the main resonance the coherent sum exceeds any single-channel contribution, demonstrating constructive superposition, while in the valley between the two dominant peaks the coherent signal drops below the individual curves, revealing destructive interference.

The presence of the applied magnetic field introduces an asymmetry between the positive- and negative-$m$ channels. In the absence of the field, the $+m$ and $-m$ pathways would be exactly degenerate due to time-reversal symmetry. For the finite field used in Fig.~\ref{fig:channel_decomp}, however, the orbital magnetic splitting separates the resonance positions of the two families, producing a characteristic doublet structure that is visible in the decomposition. The spiral dislocation further modulates this asymmetry by modifying the radial overlaps differently for different angular-momentum channels. This interplay between geometric and magnetic symmetry breaking is a distinctive feature of the model and would not be apparent from the total spectrum alone.

\begin{figure}[tbhp]
\centering
\includegraphics[width=\linewidth]{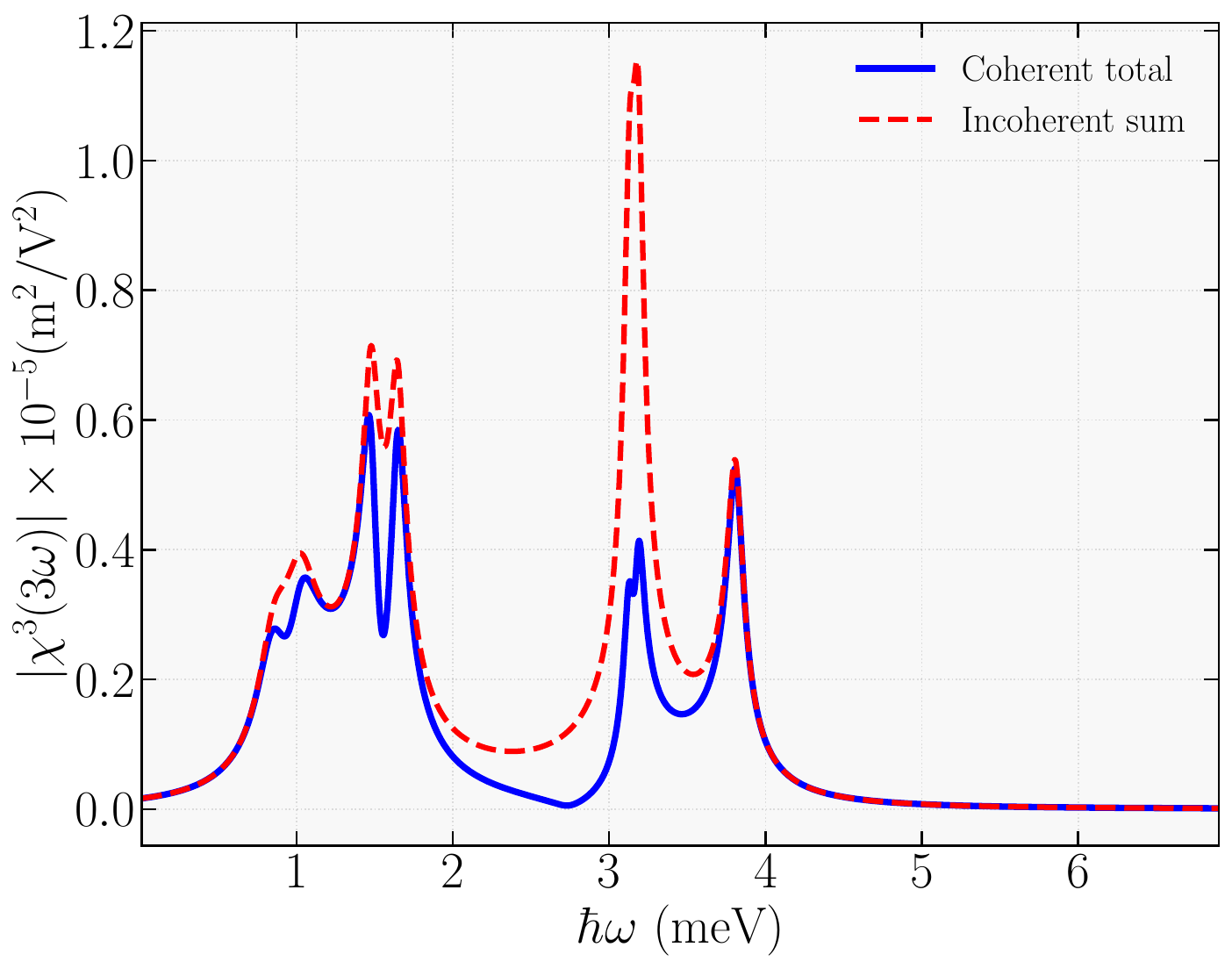}
\caption{Comparison of the coherent total (solid line) with the incoherent sum of all channel contributions (dashed line) for the third-harmonic generation susceptibility at $\beta=3.0\,\mathrm{nm}$, $B=0.1\,\mathrm{T}$ and $\Gamma=0.10\,\mathrm{meV}$.}
\label{fig:coherent_vs_incoherent}
\end{figure}

Figure~\ref{fig:coherent_vs_incoherent} provides a complementary diagnostic by comparing the coherent total $|\sum_P\chi_P^{(3)}|$ with the incoherent sum $\sum_P|\chi_P^{(3)}|$. The incoherent sum represents the hypothetical signal that would be observed if all pathway contributions were added in intensity rather than in amplitude. Any difference between the two curves is therefore a direct measure of interference. The comparison reveals that across most of the spectral range, the coherent total lies below the incoherent sum, indicating that destructive interference is the prevailing mechanism. This is physically sensible: the different pathways acquire different complex phases from their respective energy denominators, and because those denominators involve combinations of one-, two-, and three-photon detunings, the relative phases change rapidly across the spectrum. The net effect is a partial cancellation that reduces the total signal below the sum of individual intensities, except in narrow windows around the dominant resonance where the phases happen to align constructively. This interference-driven spectral reshaping is one of the mechanisms through which the spiral dislocation exerts fine control over the THG profile: by modifying the radial overlaps and energy spacings in a channel-dependent manner, $\beta$ changes not only the individual pathway weights but also the conditions under which constructive or destructive superposition occurs.

\subsection{General discussion}
\label{sec:discussion}

The analysis supports three central conclusions. First, spiral dislocation primarily changes the nonlinear response by reshaping the radial confinement and redistributing dipole strength among allowed transitions. The defect, therefore, acts as a genuinely geometric control parameter rather than as a simple perturbative energy shift. Second, the model's axial symmetry remains intact even in the torsion-deformed geometry, which explains why SHG is forbidden while THG persists through multistep dipole-allowed chains. Third, the magnetic field and the spiral dislocation act cooperatively: the field breaks the degeneracy between positive- and negative-angular-momentum channels, while the defect changes their radial overlaps and detunings in a state-dependent way.

The three extensions introduced above add considerable physical depth to these conclusions. The dephasing analysis (Extension~I) demonstrates that the THG signal amplitude scales approximately as $\Gamma^{-3}$ in the resolved-peak regime, making it a substantially more sensitive probe of decoherence than linear absorption or lower-order optical responses. The 3D waterfall spectra (Extension~II) condense the combined influence of geometry and magnetic field into comprehensive 3D visualizations that identify the operating regimes for enhanced THG intensity and target emission energy. The channel decomposition (Extension~III) reveals that the total THG signal is shaped not only by the individual pathway strengths but also by inter-channel interference, which is predominantly destructive across most of the spectral range but becomes constructive near selected resonances. This interference mechanism provides an additional degree of control: by tuning $\beta$ or $B$, one can modify the relative phases of the contributing pathways, thereby enhancing or suppressing the net signal in specific spectral windows.

\section{Conclusions}
\label{sec:conclusion}

We have presented a comprehensive and consolidated analysis of nonlinear optical generation in a mesoscopic quantum ring with a spiral dislocation. Starting from the Schr\"odinger equation in a torsion-deformed two-dimensional background, we derived the effective radial problem, analyzed the bound states and dipole-allowed transitions, and used them to construct the nonlinear susceptibilities. The geometric defect modifies the effective radial landscape and redistributes dipole strengths. Crucially, we demonstrated that the topological defect and the magnetic field act cooperatively to drive a state-dependent reorganization of the spectrum---simultaneously pushing the fundamental transition toward lower energies (red-shift) and the higher-energy multiphoton resonances toward higher values (blue-shift). At the same time, the model preserves axial symmetry and the electric-dipole selection rule $\Delta m=\pm 1$, which strictly forbids SHG in the dipole approximation but leaves THG allowed through multistep transition chains.

The dephasing-resolved study quantified how the peak amplitude, resonance position, and linewidth of the THG features evolve as a function of the dephasing broadening. It revealed the approximate $\Gamma^{-3}$ sensitivity of the peak height in the resolved-resonance regime, establishing THG as a sharp probe of geometry-induced spectral changes. The 3D waterfall spectra condensed the influence of dislocation strength and magnetic field into comprehensive 3D visualizations, revealing operating regions in which the THG intensity can be enhanced or shifted spectrally. The channel-resolved decomposition identified the dominant symmetry-allowed transition chains and demonstrated that inter-channel interference plays a central role in shaping the total spectral profile, providing a microscopic mechanism through which the spiral dislocation exerts fine spectral control. Overall, our results reinforce the idea that spiral dislocation is not merely a passive structural flaw, but a robust topological tool for tailoring nonlinear light--matter interactions in mesoscopic ring-shaped nanostructures, guiding the design of geometry-assisted optoelectronic control strategies.

\section*{Acknowledgement}

This work was partially supported by the Brazilian agencies CAPES, CNPq, FAPES, and FAPEMA. The authors acknowledge support from CNPq/306308/2022-3, FAPEMA/UNIVERSAL-06395/22, and FAPEMA/APP-12256/22. This study was financed in part by the Coordena\c{c}\~ao de Aperfei\c{c}oamento de Pessoal de N\'ivel Superior - Brasil (CAPES) - Finance Code 001. Assafrão is also grateful for the computational resources of the Sci-com/NC3/UFES.

\section*{Data Availability Statement}

The data that support the findings of this study are available from the corresponding author upon reasonable request.


\begin{thebibliography}{59}%
	\makeatletter
	\providecommand \@ifxundefined [1]{%
		\@ifx{#1\undefined}
	}%
	\providecommand \@ifnum [1]{%
		\ifnum #1\expandafter \@firstoftwo
		\else \expandafter \@secondoftwo
		\fi
	}%
	\providecommand \@ifx [1]{%
		\ifx #1\expandafter \@firstoftwo
		\else \expandafter \@secondoftwo
		\fi
	}%
	\providecommand \natexlab [1]{#1}%
	\providecommand \enquote  [1]{``#1''}%
	\providecommand \bibnamefont  [1]{#1}%
	\providecommand \bibfnamefont [1]{#1}%
	\providecommand \citenamefont [1]{#1}%
	\providecommand \href@noop [0]{\@secondoftwo}%
	\providecommand \href [0]{\begingroup \@sanitize@url \@href}%
	\providecommand \@href[1]{\@@startlink{#1}\@@href}%
	\providecommand \@@href[1]{\endgroup#1\@@endlink}%
	\providecommand \@sanitize@url [0]{\catcode `\\12\catcode `\$12\catcode
		`\&12\catcode `\#12\catcode `\^12\catcode `\_12\catcode `\%12\relax}%
	\providecommand \@@startlink[1]{}%
	\providecommand \@@endlink[0]{}%
	\providecommand \url  [0]{\begingroup\@sanitize@url \@url }%
	\providecommand \@url [1]{\endgroup\@href {#1}{\urlprefix }}%
	\providecommand \urlprefix  [0]{URL }%
	\providecommand \Eprint [0]{\href }%
	\providecommand \doibase [0]{https://doi.org/}%
	\providecommand \selectlanguage [0]{\@gobble}%
	\providecommand \bibinfo  [0]{\@secondoftwo}%
	\providecommand \bibfield  [0]{\@secondoftwo}%
	\providecommand \translation [1]{[#1]}%
	\providecommand \BibitemOpen [0]{}%
	\providecommand \bibitemStop [0]{}%
	\providecommand \bibitemNoStop [0]{.\EOS\space}%
	\providecommand \EOS [0]{\spacefactor3000\relax}%
	\providecommand \BibitemShut  [1]{\csname bibitem#1\endcsname}%
	\let\auto@bib@innerbib\@empty
	%</preamble>
	\bibitem [{\citenamefont {Aharonov}\ and\ \citenamefont
		{Bohm}(1959)}]{Aharonov1959}%
	\BibitemOpen
	\bibfield  {author} {\bibinfo {author} {\bibfnamefont {Y.}~\bibnamefont
			{Aharonov}}\ and\ \bibinfo {author} {\bibfnamefont {D.}~\bibnamefont
			{Bohm}},\ }\bibfield  {title} {\bibinfo {title} {Significance of
			electromagnetic potentials in the quantum theory},\ }\href
	{https://doi.org/10.1103/PhysRev.115.485} {\bibfield  {journal} {\bibinfo
			{journal} {Physical Review}\ }\textbf {\bibinfo {volume} {115}},\ \bibinfo
		{pages} {485} (\bibinfo {year} {1959})}\BibitemShut {NoStop}%
	\bibitem [{\citenamefont {Lorke}\ \emph {et~al.}(2000)\citenamefont {Lorke},
		\citenamefont {Luyken}, \citenamefont {Govorov}, \citenamefont {Kotthaus},
		\citenamefont {Garcia},\ and\ \citenamefont {Petroff}}]{Lorke2000}%
	\BibitemOpen
	\bibfield  {author} {\bibinfo {author} {\bibfnamefont {A.}~\bibnamefont
			{Lorke}}, \bibinfo {author} {\bibfnamefont {R.~J.}\ \bibnamefont {Luyken}},
		\bibinfo {author} {\bibfnamefont {A.~O.}\ \bibnamefont {Govorov}}, \bibinfo
		{author} {\bibfnamefont {J.~M.}\ \bibnamefont {Kotthaus}}, \bibinfo {author}
		{\bibfnamefont {J.~M.}\ \bibnamefont {Garcia}},\ and\ \bibinfo {author}
		{\bibfnamefont {P.~M.}\ \bibnamefont {Petroff}},\ }\bibfield  {title}
	{\bibinfo {title} {Spectroscopy of nanoscopic semiconductor rings},\ }\href
	{https://doi.org/10.1103/PhysRevLett.84.2223} {\bibfield  {journal} {\bibinfo
			{journal} {Physical Review Letters}\ }\textbf {\bibinfo {volume} {84}},\
		\bibinfo {pages} {2223} (\bibinfo {year} {2000})}\BibitemShut {NoStop}%
	\bibitem [{\citenamefont {Fuhrer}\ \emph {et~al.}(2001)\citenamefont {Fuhrer},
		\citenamefont {L"uscher}, \citenamefont {Ihn}, \citenamefont {Heinzel},
		\citenamefont {Ensslin}, \citenamefont {Wegscheider},\ and\ \citenamefont
		{Bichler}}]{Fuhrer2001}%
	\BibitemOpen
	\bibfield  {author} {\bibinfo {author} {\bibfnamefont {A.}~\bibnamefont
			{Fuhrer}}, \bibinfo {author} {\bibfnamefont {S.}~\bibnamefont {L"uscher}},
		\bibinfo {author} {\bibfnamefont {T.}~\bibnamefont {Ihn}}, \bibinfo {author}
		{\bibfnamefont {T.}~\bibnamefont {Heinzel}}, \bibinfo {author} {\bibfnamefont
			{K.}~\bibnamefont {Ensslin}}, \bibinfo {author} {\bibfnamefont
			{W.}~\bibnamefont {Wegscheider}},\ and\ \bibinfo {author} {\bibfnamefont
			{M.}~\bibnamefont {Bichler}},\ }\bibfield  {title} {\bibinfo {title} {Energy
			spectra of quantum rings},\ }\href {https://doi.org/10.1038/35101552}
	{\bibfield  {journal} {\bibinfo  {journal} {Nature}\ }\textbf {\bibinfo
			{volume} {413}},\ \bibinfo {pages} {822} (\bibinfo {year}
		{2001})}\BibitemShut {NoStop}%
	\bibitem [{\citenamefont {Olariu}\ and\ \citenamefont
		{Popescu}(1985)}]{RevModPhys.57.339}%
	\BibitemOpen
	\bibfield  {author} {\bibinfo {author} {\bibfnamefont {S.}~\bibnamefont
			{Olariu}}\ and\ \bibinfo {author} {\bibfnamefont {I.~I.}\ \bibnamefont
			{Popescu}},\ }\bibfield  {title} {\bibinfo {title} {The quantum effects of
			electromagnetic fluxes},\ }\href {https://doi.org/10.1103/RevModPhys.57.339}
	{\bibfield  {journal} {\bibinfo  {journal} {Rev. Mod. Phys.}\ }\textbf
		{\bibinfo {volume} {57}},\ \bibinfo {pages} {339} (\bibinfo {year}
		{1985})}\BibitemShut {NoStop}%
	\bibitem [{\citenamefont {Chandrasekhar}\ \emph {et~al.}(1985)\citenamefont
		{Chandrasekhar}, \citenamefont {Rooks}, \citenamefont {Wind},\ and\
		\citenamefont {Prober}}]{PRL.1985.55.1610}%
	\BibitemOpen
	\bibfield  {author} {\bibinfo {author} {\bibfnamefont {V.}~\bibnamefont
			{Chandrasekhar}}, \bibinfo {author} {\bibfnamefont {M.~J.}\ \bibnamefont
			{Rooks}}, \bibinfo {author} {\bibfnamefont {S.}~\bibnamefont {Wind}},\ and\
		\bibinfo {author} {\bibfnamefont {D.~E.}\ \bibnamefont {Prober}},\ }\bibfield
	{title} {\bibinfo {title} {Observation of aharonov-bohm electron
			interference effects with periods $\frac{h}{e}$ and $\frac{h}{2e}$ in
			individual micron-size, normal-metal rings},\ }\href
	{https://doi.org/10.1103/PhysRevLett.55.1610} {\bibfield  {journal} {\bibinfo
			{journal} {Phys. Rev. Lett.}\ }\textbf {\bibinfo {volume} {55}},\ \bibinfo
		{pages} {1610} (\bibinfo {year} {1985})}\BibitemShut {NoStop}%
	\bibitem [{\citenamefont {Mailly}\ \emph {et~al.}(1993)\citenamefont {Mailly},
		\citenamefont {Chapelier},\ and\ \citenamefont {Benoit}}]{PRL.1993.70.2020}%
	\BibitemOpen
	\bibfield  {author} {\bibinfo {author} {\bibfnamefont {D.}~\bibnamefont
			{Mailly}}, \bibinfo {author} {\bibfnamefont {C.}~\bibnamefont {Chapelier}},\
		and\ \bibinfo {author} {\bibfnamefont {A.}~\bibnamefont {Benoit}},\
	}\bibfield  {title} {\bibinfo {title} {Experimental observation of persistent
			currents in gaas-algaas single loop},\ }\href
	{https://doi.org/10.1103/PhysRevLett.70.2020} {\bibfield  {journal} {\bibinfo
			{journal} {Phys. Rev. Lett.}\ }\textbf {\bibinfo {volume} {70}},\ \bibinfo
		{pages} {2020} (\bibinfo {year} {1993})}\BibitemShut {NoStop}%
	\bibitem [{\citenamefont {Pershin}\ and\ \citenamefont
		{Piermarocchi}(2005)}]{PhysRevB.72.125348}%
	\BibitemOpen
	\bibfield  {author} {\bibinfo {author} {\bibfnamefont {Y.~V.}\ \bibnamefont
			{Pershin}}\ and\ \bibinfo {author} {\bibfnamefont {C.}~\bibnamefont
			{Piermarocchi}},\ }\bibfield  {title} {\bibinfo {title} {Persistent and
			radiation-induced currents in distorted quantum rings},\ }\href
	{https://doi.org/10.1103/PhysRevB.72.125348} {\bibfield  {journal} {\bibinfo
			{journal} {Phys. Rev. B}\ }\textbf {\bibinfo {volume} {72}},\ \bibinfo
		{pages} {125348} (\bibinfo {year} {2005})}\BibitemShut {NoStop}%
	\bibitem [{\citenamefont {Zipper}\ \emph {et~al.}(2006)\citenamefont {Zipper},
		\citenamefont {Kurpas}, \citenamefont {Szelag}, \citenamefont {Dajka},\ and\
		\citenamefont {Szopa}}]{PhysRevB.74.125426}%
	\BibitemOpen
	\bibfield  {author} {\bibinfo {author} {\bibfnamefont {E.}~\bibnamefont
			{Zipper}}, \bibinfo {author} {\bibfnamefont {M.}~\bibnamefont {Kurpas}},
		\bibinfo {author} {\bibfnamefont {M.}~\bibnamefont {Szelag}}, \bibinfo
		{author} {\bibfnamefont {J.}~\bibnamefont {Dajka}},\ and\ \bibinfo {author}
		{\bibfnamefont {M.}~\bibnamefont {Szopa}},\ }\bibfield  {title} {\bibinfo
		{title} {Flux qubit on a mesoscopic nonsuperconducting ring},\ }\href
	{https://doi.org/10.1103/PhysRevB.74.125426} {\bibfield  {journal} {\bibinfo
			{journal} {Phys. Rev. B}\ }\textbf {\bibinfo {volume} {74}},\ \bibinfo
		{pages} {125426} (\bibinfo {year} {2006})}\BibitemShut {NoStop}%
	\bibitem [{\citenamefont {Tan}\ and\ \citenamefont
		{Inkson}(1999)}]{TanInkson1999}%
	\BibitemOpen
	\bibfield  {author} {\bibinfo {author} {\bibfnamefont {W.-C.}\ \bibnamefont
			{Tan}}\ and\ \bibinfo {author} {\bibfnamefont {J.~C.}\ \bibnamefont
			{Inkson}},\ }\bibfield  {title} {\bibinfo {title} {Magnetization, persistent
			currents, and their relation in quantum rings and dots},\ }\href
	{https://doi.org/10.1103/PhysRevB.60.5626} {\bibfield  {journal} {\bibinfo
			{journal} {Physical Review B}\ }\textbf {\bibinfo {volume} {60}},\ \bibinfo
		{pages} {5626} (\bibinfo {year} {1999})}\BibitemShut {NoStop}%
	\bibitem [{\citenamefont {Aichinger}\ \emph {et~al.}(2006)\citenamefont
		{Aichinger}, \citenamefont {Chin}, \citenamefont {Krotscheck},\ and\
		\citenamefont {R\"as\"anen}}]{PRB.2006.73.195310}%
	\BibitemOpen
	\bibfield  {author} {\bibinfo {author} {\bibfnamefont {M.}~\bibnamefont
			{Aichinger}}, \bibinfo {author} {\bibfnamefont {S.~A.}\ \bibnamefont {Chin}},
		\bibinfo {author} {\bibfnamefont {E.}~\bibnamefont {Krotscheck}},\ and\
		\bibinfo {author} {\bibfnamefont {E.}~\bibnamefont {R\"as\"anen}},\
	}\bibfield  {title} {\bibinfo {title} {Effects of geometry and impurities on
			quantum rings in magnetic fields},\ }\href
	{https://doi.org/10.1103/PhysRevB.73.195310} {\bibfield  {journal} {\bibinfo
			{journal} {Phys. Rev. B}\ }\textbf {\bibinfo {volume} {73}},\ \bibinfo
		{pages} {195310} (\bibinfo {year} {2006})}\BibitemShut {NoStop}%
	\bibitem [{\citenamefont {Bruno-Alfonso}\ and\ \citenamefont
		{Latg\'e}(2008)}]{PhysRevB.77.205303}%
	\BibitemOpen
	\bibfield  {author} {\bibinfo {author} {\bibfnamefont {A.}~\bibnamefont
			{Bruno-Alfonso}}\ and\ \bibinfo {author} {\bibfnamefont {A.}~\bibnamefont
			{Latg\'e}},\ }\bibfield  {title} {\bibinfo {title} {Quantum rings of
			arbitrary shape and non-uniform width in a threading magnetic field},\ }\href
	{https://doi.org/10.1103/PhysRevB.77.205303} {\bibfield  {journal} {\bibinfo
			{journal} {Phys. Rev. B}\ }\textbf {\bibinfo {volume} {77}},\ \bibinfo
		{pages} {205303} (\bibinfo {year} {2008})}\BibitemShut {NoStop}%
	\bibitem [{\citenamefont {Bandos}\ \emph {et~al.}(2006)\citenamefont {Bandos},
		\citenamefont {Cantarero},\ and\ \citenamefont
		{Garc{\'{\i}}a-Crist{\'{o}}bal}}]{EPJB.2006.53.99}%
	\BibitemOpen
	\bibfield  {author} {\bibinfo {author} {\bibfnamefont {T.~V.}\ \bibnamefont
			{Bandos}}, \bibinfo {author} {\bibfnamefont {A.}~\bibnamefont {Cantarero}},\
		and\ \bibinfo {author} {\bibfnamefont {A.}~\bibnamefont
			{Garc{\'{\i}}a-Crist{\'{o}}bal}},\ }\bibfield  {title} {\bibinfo {title}
		{Finite size effects on the optical transitions in quantum rings under a
			magnetic field},\ }\href {https://doi.org/10.1140/epjb/e2006-00351-2}
	{\bibfield  {journal} {\bibinfo  {journal} {The European Physical Journal B}\
		}\textbf {\bibinfo {volume} {53}},\ \bibinfo {pages} {99} (\bibinfo {year}
		{2006})}\BibitemShut {NoStop}%
	\bibitem [{\citenamefont {Liang}\ \emph {et~al.}(2011)\citenamefont {Liang},
		\citenamefont {Xie},\ and\ \citenamefont {Shen}}]{Liang2011}%
	\BibitemOpen
	\bibfield  {author} {\bibinfo {author} {\bibfnamefont {S.}~\bibnamefont
			{Liang}}, \bibinfo {author} {\bibfnamefont {W.}~\bibnamefont {Xie}},\ and\
		\bibinfo {author} {\bibfnamefont {H.}~\bibnamefont {Shen}},\ }\bibfield
	{title} {\bibinfo {title} {Optical properties in a two-dimensional quantum
			ring: Confinement potential and aharonov--bohm effect},\ }\href
	{https://doi.org/10.1016/j.optcom.2011.08.080} {\bibfield  {journal}
		{\bibinfo  {journal} {Optics Communications}\ }\textbf {\bibinfo {volume}
			{284}},\ \bibinfo {pages} {5818} (\bibinfo {year} {2011})}\BibitemShut
	{NoStop}%
	\bibitem [{\citenamefont {Khajeh~Salehani}(2023)}]{Salehani2023}%
	\BibitemOpen
	\bibfield  {author} {\bibinfo {author} {\bibfnamefont {H.}~\bibnamefont
			{Khajeh~Salehani}},\ }\bibfield  {title} {\bibinfo {title} {Optical
			absorption in concentric double quantum rings},\ }\href
	{https://doi.org/10.1007/s11082-023-04939-x} {\bibfield  {journal} {\bibinfo
			{journal} {Optical and Quantum Electronics}\ }\textbf {\bibinfo {volume}
			{55}},\ \bibinfo {pages} {644} (\bibinfo {year} {2023})}\BibitemShut
	{NoStop}%
	\bibitem [{\citenamefont {Chang}(2023)}]{Chang2023}%
	\BibitemOpen
	\bibfield  {author} {\bibinfo {author} {\bibfnamefont {C.}~\bibnamefont
			{Chang}},\ }\bibfield  {title} {\bibinfo {title} {Studies on the
			third-harmonic generations in a quantum ring with magnetic field},\ }\href
	{https://doi.org/10.1140/epjp/s13360-023-03733-8} {\bibfield  {journal}
		{\bibinfo  {journal} {The European Physical Journal Plus}\ }\textbf {\bibinfo
			{volume} {138}},\ \bibinfo {pages} {116} (\bibinfo {year}
		{2023})}\BibitemShut {NoStop}%
	\bibitem [{\citenamefont {Lima}\ \emph {et~al.}(2023)\citenamefont {Lima},
		\citenamefont {Azevedo}, \citenamefont {Pereira}, \citenamefont
		{Filgueiras},\ and\ \citenamefont {Silva}}]{Lima2023}%
	\BibitemOpen
	\bibfield  {author} {\bibinfo {author} {\bibfnamefont {D.~F.}\ \bibnamefont
			{Lima}}, \bibinfo {author} {\bibfnamefont {F.~d.~S.}\ \bibnamefont
			{Azevedo}}, \bibinfo {author} {\bibfnamefont {L.~F.~C.}\ \bibnamefont
			{Pereira}}, \bibinfo {author} {\bibfnamefont {C.}~\bibnamefont
			{Filgueiras}},\ and\ \bibinfo {author} {\bibfnamefont {E.~O.}\ \bibnamefont
			{Silva}},\ }\bibfield  {title} {\bibinfo {title} {Optical and electronic
			properties of a two-dimensional quantum ring under rotating effects},\ }\href
	{https://doi.org/10.1016/j.aop.2023.169547} {\bibfield  {journal} {\bibinfo
			{journal} {Annals of Physics}\ }\textbf {\bibinfo {volume} {459}},\ \bibinfo
		{pages} {169547} (\bibinfo {year} {2023})}\BibitemShut {NoStop}%
	\bibitem [{\citenamefont {Bejan}\ and\ \citenamefont {Stan}(2025)}]{Bejan2025}%
	\BibitemOpen
	\bibfield  {author} {\bibinfo {author} {\bibfnamefont {D.}~\bibnamefont
			{Bejan}}\ and\ \bibinfo {author} {\bibfnamefont {C.}~\bibnamefont {Stan}},\
	}\bibfield  {title} {\bibinfo {title} {Nonlinear optical rectification,
			second and third harmonic generation in quantum dot-double quantum rings:
			electric field and geometry effects},\ }\href
	{https://doi.org/10.1007/s11082-025-08403-w} {\bibfield  {journal} {\bibinfo
			{journal} {Optical and Quantum Electronics}\ }\textbf {\bibinfo {volume}
			{57}},\ \bibinfo {pages} {476} (\bibinfo {year} {2025})}\BibitemShut
	{NoStop}%
	\bibitem [{\citenamefont {Duque}\ \emph {et~al.}(2013)\citenamefont {Duque},
		\citenamefont {Morales}, \citenamefont {Mora-Ramos},\ and\ \citenamefont
		{Duque}}]{Duque2010}%
	\BibitemOpen
	\bibfield  {author} {\bibinfo {author} {\bibfnamefont {C.}~\bibnamefont
			{Duque}}, \bibinfo {author} {\bibfnamefont {A.}~\bibnamefont {Morales}},
		\bibinfo {author} {\bibfnamefont {M.}~\bibnamefont {Mora-Ramos}},\ and\
		\bibinfo {author} {\bibfnamefont {C.}~\bibnamefont {Duque}},\ }\bibfield
	{title} {\bibinfo {title} {Optical nonlinearities associated to applied
			electric fields in parabolic two-dimensional quantum rings},\ }\href
	{https://doi.org/https://doi.org/10.1016/j.jlumin.2013.04.039} {\bibfield
		{journal} {\bibinfo  {journal} {Journal of Luminescence}\ }\textbf {\bibinfo
			{volume} {143}},\ \bibinfo {pages} {81} (\bibinfo {year} {2013})}\BibitemShut
	{NoStop}%
	\bibitem [{\citenamefont {Duque}\ \emph {et~al.}(2012)\citenamefont {Duque},
		\citenamefont {Mora-Ramos},\ and\ \citenamefont {Duque}}]{AdP.2012.524.327}%
	\BibitemOpen
	\bibfield  {author} {\bibinfo {author} {\bibfnamefont {C.}~\bibnamefont
			{Duque}}, \bibinfo {author} {\bibfnamefont {M.}~\bibnamefont {Mora-Ramos}},\
		and\ \bibinfo {author} {\bibfnamefont {C.}~\bibnamefont {Duque}},\ }\bibfield
	{title} {\bibinfo {title} {Quantum disc plus inverse square potential. an
			analytical model for two-dimensional quantum rings: Study of nonlinear
			optical properties},\ }\href
	{https://doi.org/https://doi.org/10.1002/andp.201200055} {\bibfield
		{journal} {\bibinfo  {journal} {Annalen der Physik}\ }\textbf {\bibinfo
			{volume} {524}},\ \bibinfo {pages} {327} (\bibinfo {year}
		{2012})}\BibitemShut {NoStop}%
	\bibitem [{\citenamefont {Gumber}\ \emph {et~al.}(2016)\citenamefont {Gumber},
		\citenamefont {Gambhir}, \citenamefont {Jha},\ and\ \citenamefont
		{Mohan}}]{rashba}%
	\BibitemOpen
	\bibfield  {author} {\bibinfo {author} {\bibfnamefont {S.}~\bibnamefont
			{Gumber}}, \bibinfo {author} {\bibfnamefont {M.}~\bibnamefont {Gambhir}},
		\bibinfo {author} {\bibfnamefont {P.~K.}\ \bibnamefont {Jha}},\ and\ \bibinfo
		{author} {\bibfnamefont {M.}~\bibnamefont {Mohan}},\ }\bibfield  {title}
	{\bibinfo {title} {{Optical response of a two dimensional quantum ring in
				presence of Rashba spin orbit coupling}},\ }\href
	{https://doi.org/10.1063/1.4942015} {\bibfield  {journal} {\bibinfo
			{journal} {Journal of Applied Physics}\ }\textbf {\bibinfo {volume} {119}},\
		\bibinfo {pages} {073101} (\bibinfo {year} {2016})}\BibitemShut {NoStop}%
	\bibitem [{\citenamefont {Nasri}(2021)}]{ringparallel}%
	\BibitemOpen
	\bibfield  {author} {\bibinfo {author} {\bibfnamefont {D.}~\bibnamefont
			{Nasri}},\ }\bibfield  {title} {\bibinfo {title} {Electronic and optical
			properties of eccentric quantum ring under parallel magnetic field},\ }\href
	{https://doi.org/https://doi.org/10.1016/j.physb.2021.413077} {\bibfield
		{journal} {\bibinfo  {journal} {Physica B: Condensed Matter}\ }\textbf
		{\bibinfo {volume} {615}},\ \bibinfo {pages} {413077} (\bibinfo {year}
		{2021})}\BibitemShut {NoStop}%
	\bibitem [{\citenamefont {Vinasco}\ \emph
		{et~al.}(2019{\natexlab{a}})\citenamefont {Vinasco}, \citenamefont {Radu},
		\citenamefont {Restrepo}, \citenamefont {Morales}, \citenamefont
		{Mora-Ramos},\ and\ \citenamefont {Duque}}]{OM.2019.91.309}%
	\BibitemOpen
	\bibfield  {author} {\bibinfo {author} {\bibfnamefont {J.}~\bibnamefont
			{Vinasco}}, \bibinfo {author} {\bibfnamefont {A.}~\bibnamefont {Radu}},
		\bibinfo {author} {\bibfnamefont {R.}~\bibnamefont {Restrepo}}, \bibinfo
		{author} {\bibfnamefont {A.}~\bibnamefont {Morales}}, \bibinfo {author}
		{\bibfnamefont {M.}~\bibnamefont {Mora-Ramos}},\ and\ \bibinfo {author}
		{\bibfnamefont {C.}~\bibnamefont {Duque}},\ }\bibfield  {title} {\bibinfo
		{title} {Magnetic field effects on intraband transitions in elliptically
			polarized laser-dressed quantum rings},\ }\href
	{https://doi.org/https://doi.org/10.1016/j.optmat.2019.03.016} {\bibfield
		{journal} {\bibinfo  {journal} {Optical Materials}\ }\textbf {\bibinfo
			{volume} {91}},\ \bibinfo {pages} {309} (\bibinfo {year}
		{2019}{\natexlab{a}})}\BibitemShut {NoStop}%
	\bibitem [{\citenamefont {Vinasco}\ \emph
		{et~al.}(2019{\natexlab{b}})\citenamefont {Vinasco}, \citenamefont {Radu},
		\citenamefont {Niculescu}, \citenamefont {Mora-Ramos}, \citenamefont {Feddi},
		\citenamefont {Tulupenko}, \citenamefont {Restrepo}, \citenamefont
		{Kasapoglu}, \citenamefont {Morales},\ and\ \citenamefont
		{Duque}}]{SR.2019.9.1427}%
	\BibitemOpen
	\bibfield  {author} {\bibinfo {author} {\bibfnamefont {J.~A.}\ \bibnamefont
			{Vinasco}}, \bibinfo {author} {\bibfnamefont {A.}~\bibnamefont {Radu}},
		\bibinfo {author} {\bibfnamefont {E.}~\bibnamefont {Niculescu}}, \bibinfo
		{author} {\bibfnamefont {M.~E.}\ \bibnamefont {Mora-Ramos}}, \bibinfo
		{author} {\bibfnamefont {E.}~\bibnamefont {Feddi}}, \bibinfo {author}
		{\bibfnamefont {V.}~\bibnamefont {Tulupenko}}, \bibinfo {author}
		{\bibfnamefont {R.~L.}\ \bibnamefont {Restrepo}}, \bibinfo {author}
		{\bibfnamefont {E.}~\bibnamefont {Kasapoglu}}, \bibinfo {author}
		{\bibfnamefont {A.~L.}\ \bibnamefont {Morales}},\ and\ \bibinfo {author}
		{\bibfnamefont {C.~A.}\ \bibnamefont {Duque}},\ }\bibfield  {title} {\bibinfo
		{title} {Electronic states in gaas-(al,ga)as eccentric quantum rings under
			nonresonant intense laser and magnetic fields},\ }\href
	{https://doi.org/10.1038/s41598-018-38114-0} {\bibfield  {journal} {\bibinfo
			{journal} {Scientific Reports}\ }\textbf {\bibinfo {volume} {9}},\ \bibinfo
		{pages} {1427} (\bibinfo {year} {2019}{\natexlab{b}})}\BibitemShut {NoStop}%
	\bibitem [{\citenamefont {Xie}(2013)}]{SM.2013.58.94}%
	\BibitemOpen
	\bibfield  {author} {\bibinfo {author} {\bibfnamefont {W.}~\bibnamefont
			{Xie}},\ }\bibfield  {title} {\bibinfo {title} {Photoionization cross section
			in a two-dimensional quantum ring: Aharonov–bohm effect},\ }\href
	{https://doi.org/https://doi.org/10.1016/j.spmi.2013.03.010} {\bibfield
		{journal} {\bibinfo  {journal} {Superlattices and Microstructures}\ }\textbf
		{\bibinfo {volume} {58}},\ \bibinfo {pages} {94} (\bibinfo {year}
		{2013})}\BibitemShut {NoStop}%
	\bibitem [{\citenamefont {Pereira}\ \emph
		{et~al.}(2024{\natexlab{a}})\citenamefont {Pereira}, \citenamefont {dos
			S~Azevedo}, \citenamefont {Pereira},\ and\ \citenamefont
		{Silva}}]{CTP.2024.76.105701}%
	\BibitemOpen
	\bibfield  {author} {\bibinfo {author} {\bibfnamefont {C.~M.~O.}\
			\bibnamefont {Pereira}}, \bibinfo {author} {\bibfnamefont {F.}~\bibnamefont
			{dos S~Azevedo}}, \bibinfo {author} {\bibfnamefont {L.~F.~C.}\ \bibnamefont
			{Pereira}},\ and\ \bibinfo {author} {\bibfnamefont {E.~O.}\ \bibnamefont
			{Silva}},\ }\bibfield  {title} {\bibinfo {title} {Rotating effects on the
			photoionization cross-section of a 2d quantum ring},\ }\href
	{https://doi.org/10.1088/1572-9494/ad597c} {\bibfield  {journal} {\bibinfo
			{journal} {Communications in Theoretical Physics}\ }\textbf {\bibinfo
			{volume} {76}},\ \bibinfo {pages} {105701} (\bibinfo {year}
		{2024}{\natexlab{a}})}\BibitemShut {NoStop}%
	\bibitem [{\citenamefont {Pereira}\ \emph
		{et~al.}(2024{\natexlab{b}})\citenamefont {Pereira}, \citenamefont
		{Azevedo},\ and\ \citenamefont {Silva}}]{QR.2024.6.677}%
	\BibitemOpen
	\bibfield  {author} {\bibinfo {author} {\bibfnamefont {C.~M.~O.}\
			\bibnamefont {Pereira}}, \bibinfo {author} {\bibfnamefont {F.~d.~S.}\
			\bibnamefont {Azevedo}},\ and\ \bibinfo {author} {\bibfnamefont {E.~O.}\
			\bibnamefont {Silva}},\ }\bibfield  {title} {\bibinfo {title} {Remarks on the
			study of the electronic properties and photoionization process in rotating 2d
			quantum rings},\ }\href {https://doi.org/10.3390/quantum6040041} {\bibfield
		{journal} {\bibinfo  {journal} {Quantum Reports}\ }\textbf {\bibinfo {volume}
			{6}},\ \bibinfo {pages} {677} (\bibinfo {year}
		{2024}{\natexlab{b}})}\BibitemShut {NoStop}%
	\bibitem [{\citenamefont {Ahn}\ and\ \citenamefont {lien Chuang}(1987)}]{ahn}%
	\BibitemOpen
	\bibfield  {author} {\bibinfo {author} {\bibfnamefont {D.}~\bibnamefont
			{Ahn}}\ and\ \bibinfo {author} {\bibfnamefont {S.}~\bibnamefont {lien
				Chuang}},\ }\bibfield  {title} {\bibinfo {title} {Calculation of linear and
			nonlinear intersubband optical absorptions in a quantum well model with an
			applied electric field},\ }\href {https://doi.org/10.1109/JQE.1987.1073280}
	{\bibfield  {journal} {\bibinfo  {journal} {IEEE Journal of Quantum
				Electronics}\ }\textbf {\bibinfo {volume} {23}},\ \bibinfo {pages} {2196}
		(\bibinfo {year} {1987})}\BibitemShut {NoStop}%
	\bibitem [{\citenamefont {Rosencher}\ and\ \citenamefont
		{Bois}(1991)}]{rosencher}%
	\BibitemOpen
	\bibfield  {author} {\bibinfo {author} {\bibfnamefont {E.}~\bibnamefont
			{Rosencher}}\ and\ \bibinfo {author} {\bibfnamefont {P.}~\bibnamefont
			{Bois}},\ }\bibfield  {title} {\bibinfo {title} {Model system for optical
			nonlinearities: Asymmetric quantum wells},\ }\href
	{https://doi.org/10.1103/PhysRevB.44.11315} {\bibfield  {journal} {\bibinfo
			{journal} {Phys. Rev. B}\ }\textbf {\bibinfo {volume} {44}},\ \bibinfo
		{pages} {11315} (\bibinfo {year} {1991})}\BibitemShut {NoStop}%
	\bibitem [{\citenamefont {Li}\ \emph {et~al.}(2017)\citenamefont {Li},
		\citenamefont {Guo}, \citenamefont {Jiang},\ and\ \citenamefont
		{Hu}}]{LI2017375}%
	\BibitemOpen
	\bibfield  {author} {\bibinfo {author} {\bibfnamefont {K.}~\bibnamefont
			{Li}}, \bibinfo {author} {\bibfnamefont {K.}~\bibnamefont {Guo}}, \bibinfo
		{author} {\bibfnamefont {X.}~\bibnamefont {Jiang}},\ and\ \bibinfo {author}
		{\bibfnamefont {M.}~\bibnamefont {Hu}},\ }\bibfield  {title} {\bibinfo
		{title} {Effect of position-dependent effective mass on nonlinear optical
			properties in a quantum well},\ }\href
	{https://doi.org/https://doi.org/10.1016/j.ijleo.2016.12.011} {\bibfield
		{journal} {\bibinfo  {journal} {Optik}\ }\textbf {\bibinfo {volume} {132}},\
		\bibinfo {pages} {375} (\bibinfo {year} {2017})}\BibitemShut {NoStop}%
	\bibitem [{\citenamefont {Hu}\ \emph {et~al.}(2017)\citenamefont {Hu},
		\citenamefont {Guo}, \citenamefont {Zhang}, \citenamefont {Jiang},
		\citenamefont {Li},\ and\ \citenamefont
		{Liu}}]{doi:10.1142/S0217979217500096}%
	\BibitemOpen
	\bibfield  {author} {\bibinfo {author} {\bibfnamefont {M.}~\bibnamefont
			{Hu}}, \bibinfo {author} {\bibfnamefont {K.}~\bibnamefont {Guo}}, \bibinfo
		{author} {\bibfnamefont {Z.}~\bibnamefont {Zhang}}, \bibinfo {author}
		{\bibfnamefont {X.}~\bibnamefont {Jiang}}, \bibinfo {author} {\bibfnamefont
			{K.}~\bibnamefont {Li}},\ and\ \bibinfo {author} {\bibfnamefont
			{D.}~\bibnamefont {Liu}},\ }\bibfield  {title} {\bibinfo {title} {The effect
			of position-dependent mass on nonlinear optical absorption coefficients and
			refractive index changes in a quantum well},\ }\href
	{https://doi.org/10.1142/S0217979217500096} {\bibfield  {journal} {\bibinfo
			{journal} {International Journal of Modern Physics B}\ }\textbf {\bibinfo
			{volume} {31}},\ \bibinfo {pages} {1750009} (\bibinfo {year}
		{2017})}\BibitemShut {NoStop}%
	\bibitem [{\citenamefont {Katanaev}\ and\ \citenamefont
		{Volovich}(1992)}]{Katanaev1992}%
	\BibitemOpen
	\bibfield  {author} {\bibinfo {author} {\bibfnamefont {M.}~\bibnamefont
			{Katanaev}}\ and\ \bibinfo {author} {\bibfnamefont {I.}~\bibnamefont
			{Volovich}},\ }\bibfield  {title} {\bibinfo {title} {Theory of defects in
			solids and three-dimensional gravity},\ }\href
	{https://doi.org/https://doi.org/10.1016/0003-4916(52)90040-7} {\bibfield
		{journal} {\bibinfo  {journal} {Annals of Physics}\ }\textbf {\bibinfo
			{volume} {216}},\ \bibinfo {pages} {1} (\bibinfo {year} {1992})}\BibitemShut
	{NoStop}%
	\bibitem [{\citenamefont {Katanaev}(2005)}]{Katanaev2005}%
	\BibitemOpen
	\bibfield  {author} {\bibinfo {author} {\bibfnamefont {M.~O.}\ \bibnamefont
			{Katanaev}},\ }\bibfield  {title} {\bibinfo {title} {Geometric theory of
			defects},\ }\href {https://doi.org/10.1070/PU2005v048n07ABEH002027}
	{\bibfield  {journal} {\bibinfo  {journal} {Physics-Uspekhi}\ }\textbf
		{\bibinfo {volume} {48}},\ \bibinfo {pages} {675} (\bibinfo {year}
		{2005})}\BibitemShut {NoStop}%
	\bibitem [{\citenamefont {Harris}\ and\ \citenamefont
		{Semon}(1980)}]{FP.1980.10.151}%
	\BibitemOpen
	\bibfield  {author} {\bibinfo {author} {\bibfnamefont {J.~H.}\ \bibnamefont
			{Harris}}\ and\ \bibinfo {author} {\bibfnamefont {M.~D.}\ \bibnamefont
			{Semon}},\ }\bibfield  {title} {\bibinfo {title} {A review of the
			aharonov-carmi thought experiment concerning the inertial and electromagnetic
			vector potentials},\ }\href
	{https://doi.org/https://doi.org/10.1007/BF00709020} {\bibfield  {journal}
		{\bibinfo  {journal} {Foundations of Physics}\ }\textbf {\bibinfo {volume}
			{10}},\ \bibinfo {pages} {151} (\bibinfo {year} {1980})}\BibitemShut
	{NoStop}%
	\bibitem [{\citenamefont {Valanis}\ and\ \citenamefont
		{Panoskaltsis}(2005)}]{Valanis2005}%
	\BibitemOpen
	\bibfield  {author} {\bibinfo {author} {\bibfnamefont {K.~C.}\ \bibnamefont
			{Valanis}}\ and\ \bibinfo {author} {\bibfnamefont {V.~P.}\ \bibnamefont
			{Panoskaltsis}},\ }\bibfield  {title} {\bibinfo {title} {Material metric,
			connectivity and dislocations in continua},\ }\href
	{https://doi.org/10.1007/s00707-004-0196-9} {\bibfield  {journal} {\bibinfo
			{journal} {Acta Mechanica}\ }\textbf {\bibinfo {volume} {175}},\ \bibinfo
		{pages} {77} (\bibinfo {year} {2005})}\BibitemShut {NoStop}%
	\bibitem [{\citenamefont {Furtado}\ and\ \citenamefont
		{Moraes}(1999)}]{EPL.1999.45.279}%
	\BibitemOpen
	\bibfield  {author} {\bibinfo {author} {\bibfnamefont {C.}~\bibnamefont
			{Furtado}}\ and\ \bibinfo {author} {\bibfnamefont {F.}~\bibnamefont
			{Moraes}},\ }\bibfield  {title} {\bibinfo {title} {Landau levels in the
			presence of a screw dislocation},\ }\href
	{https://doi.org/10.1209/epl/i1999-00159-8} {\bibfield  {journal} {\bibinfo
			{journal} {Europhysics Letters}\ }\textbf {\bibinfo {volume} {45}},\ \bibinfo
		{pages} {279} (\bibinfo {year} {1999})}\BibitemShut {NoStop}%
	\bibitem [{\citenamefont {Bakke}(2014)}]{AofP.2014.346.51}%
	\BibitemOpen
	\bibfield  {author} {\bibinfo {author} {\bibfnamefont {K.}~\bibnamefont
			{Bakke}},\ }\bibfield  {title} {\bibinfo {title} {Torsion and noninertial
			effects on a nonrelativistic dirac particle},\ }\href
	{https://doi.org/https://doi.org/10.1016/j.aop.2014.04.003} {\bibfield
		{journal} {\bibinfo  {journal} {Annals of Physics}\ }\textbf {\bibinfo
			{volume} {346}},\ \bibinfo {pages} {51} (\bibinfo {year} {2014})}\BibitemShut
	{NoStop}%
	\bibitem [{\citenamefont {Bakke}\ and\ \citenamefont
		{Furtado}(2013)}]{PhysRevA.87.012130}%
	\BibitemOpen
	\bibfield  {author} {\bibinfo {author} {\bibfnamefont {K.}~\bibnamefont
			{Bakke}}\ and\ \bibinfo {author} {\bibfnamefont {C.}~\bibnamefont
			{Furtado}},\ }\bibfield  {title} {\bibinfo {title} {Abelian geometric phase
			due to the presence of an edge dislocation},\ }\href
	{https://doi.org/10.1103/PhysRevA.87.012130} {\bibfield  {journal} {\bibinfo
			{journal} {Phys. Rev. A}\ }\textbf {\bibinfo {volume} {87}},\ \bibinfo
		{pages} {012130} (\bibinfo {year} {2013})}\BibitemShut {NoStop}%
	\bibitem [{\citenamefont {Filgueiras}\ and\ \citenamefont
		{Silva}(2015)}]{PLA.2015.379.2110}%
	\BibitemOpen
	\bibfield  {author} {\bibinfo {author} {\bibfnamefont {C.}~\bibnamefont
			{Filgueiras}}\ and\ \bibinfo {author} {\bibfnamefont {E.~O.}\ \bibnamefont
			{Silva}},\ }\bibfield  {title} {\bibinfo {title} {2deg on a cylindrical shell
			with a screw dislocation},\ }\href
	{https://doi.org/https://doi.org/10.1016/j.physleta.2015.06.035} {\bibfield
		{journal} {\bibinfo  {journal} {Physics Letters A}\ }\textbf {\bibinfo
			{volume} {379}},\ \bibinfo {pages} {2110} (\bibinfo {year}
		{2015})}\BibitemShut {NoStop}%
	\bibitem [{\citenamefont {Filgueiras}\ \emph {et~al.}(2016)\citenamefont
		{Filgueiras}, \citenamefont {Rojas}, \citenamefont {Aciole},\ and\
		\citenamefont {Silva}}]{PLA.2016.380.3847}%
	\BibitemOpen
	\bibfield  {author} {\bibinfo {author} {\bibfnamefont {C.}~\bibnamefont
			{Filgueiras}}, \bibinfo {author} {\bibfnamefont {M.}~\bibnamefont {Rojas}},
		\bibinfo {author} {\bibfnamefont {G.}~\bibnamefont {Aciole}},\ and\ \bibinfo
		{author} {\bibfnamefont {E.~O.}\ \bibnamefont {Silva}},\ }\bibfield  {title}
	{\bibinfo {title} {Landau quantization, aharonov–bohm effect and
			two-dimensional pseudoharmonic quantum dot around a screw dislocation},\
	}\href {https://doi.org/https://doi.org/10.1016/j.physleta.2016.09.025}
	{\bibfield  {journal} {\bibinfo  {journal} {Physics Letters A}\ }\textbf
		{\bibinfo {volume} {380}},\ \bibinfo {pages} {3847} (\bibinfo {year}
		{2016})}\BibitemShut {NoStop}%
	\bibitem [{\citenamefont {Maia}\ and\ \citenamefont {Bakke}(2018)}]{Maia2018}%
	\BibitemOpen
	\bibfield  {author} {\bibinfo {author} {\bibfnamefont {A.~V. D.~M.}\
			\bibnamefont {Maia}}\ and\ \bibinfo {author} {\bibfnamefont {K.}~\bibnamefont
			{Bakke}},\ }\bibfield  {title} {\bibinfo {title} {Harmonic oscillator in an
			elastic medium with a spiral dislocation},\ }\href
	{https://doi.org/10.1016/j.physb.2017.12.045} {\bibfield  {journal} {\bibinfo
			{journal} {Physica B: Condensed Matter}\ }\textbf {\bibinfo {volume}
			{531}},\ \bibinfo {pages} {213} (\bibinfo {year} {2018})}\BibitemShut
	{NoStop}%
	\bibitem [{\citenamefont {da~Silva}\ and\ \citenamefont
		{Bakke}(2020)}]{daSilva2020}%
	\BibitemOpen
	\bibfield  {author} {\bibinfo {author} {\bibfnamefont {W.~C.~F.}\
			\bibnamefont {da~Silva}}\ and\ \bibinfo {author} {\bibfnamefont
			{K.}~\bibnamefont {Bakke}},\ }\bibfield  {title} {\bibinfo {title}
		{Topological effects of a spiral dislocation on quantum rings},\ }\href
	{https://doi.org/10.1016/j.aop.2020.168277} {\bibfield  {journal} {\bibinfo
			{journal} {Annals of Physics}\ }\textbf {\bibinfo {volume} {421}},\ \bibinfo
		{pages} {168277} (\bibinfo {year} {2020})}\BibitemShut {NoStop}%
	\bibitem [{\citenamefont {Hassanabadi}\ \emph {et~al.}(2026)\citenamefont
		{Hassanabadi}, \citenamefont {Guo}, \citenamefont {Lu},\ and\ \citenamefont
		{Silva}}]{Hassanabadi2026}%
	\BibitemOpen
	\bibfield  {author} {\bibinfo {author} {\bibfnamefont {H.}~\bibnamefont
			{Hassanabadi}}, \bibinfo {author} {\bibfnamefont {K.}~\bibnamefont {Guo}},
		\bibinfo {author} {\bibfnamefont {L.}~\bibnamefont {Lu}},\ and\ \bibinfo
		{author} {\bibfnamefont {E.~O.}\ \bibnamefont {Silva}},\ }\bibfield  {title}
	{\bibinfo {title} {Spiral dislocation as a tunable geometric parameter for
			optical responses in quantum rings},\ }\href
	{https://doi.org/10.1016/j.aop.2026.170346} {\bibfield  {journal} {\bibinfo
			{journal} {Annals of Physics}\ }\textbf {\bibinfo {volume} {486}},\ \bibinfo
		{pages} {170346} (\bibinfo {year} {2026})}\BibitemShut {NoStop}%
	\bibitem [{\citenamefont {Maia}\ and\ \citenamefont
		{Bakke}(2020)}]{AoP.2020.419.168229}%
	\BibitemOpen
	\bibfield  {author} {\bibinfo {author} {\bibfnamefont {A.}~\bibnamefont
			{Maia}}\ and\ \bibinfo {author} {\bibfnamefont {K.}~\bibnamefont {Bakke}},\
	}\bibfield  {title} {\bibinfo {title} {Effects of rotation on the landau
			levels in an elastic medium with a spiral dislocation},\ }\href
	{https://doi.org/https://doi.org/10.1016/j.aop.2020.168229} {\bibfield
		{journal} {\bibinfo  {journal} {Annals of Physics}\ }\textbf {\bibinfo
			{volume} {419}},\ \bibinfo {pages} {168229} (\bibinfo {year}
		{2020})}\BibitemShut {NoStop}%
	\bibitem [{\citenamefont {Maia}\ and\ \citenamefont
		{Bakke}(2019)}]{EPJC.2019.79.551}%
	\BibitemOpen
	\bibfield  {author} {\bibinfo {author} {\bibfnamefont {A.~V. D.~M.}\
			\bibnamefont {Maia}}\ and\ \bibinfo {author} {\bibfnamefont {K.}~\bibnamefont
			{Bakke}},\ }\bibfield  {title} {\bibinfo {title} {Topological and rotating
			effects on the dirac field in the spiral dislocation spacetime},\ }\href
	{https://doi.org/10.1140/epjc/s10052-019-7067-y} {\bibfield  {journal}
		{\bibinfo  {journal} {The European Physical Journal C}\ }\textbf {\bibinfo
			{volume} {79}},\ \bibinfo {pages} {551} (\bibinfo {year} {2019})}\BibitemShut
	{NoStop}%
	\bibitem [{\citenamefont {Mustafa}\ and\ \citenamefont
		{Guvendi}(2025)}]{EPJC.2025.85.34}%
	\BibitemOpen
	\bibfield  {author} {\bibinfo {author} {\bibfnamefont {O.}~\bibnamefont
			{Mustafa}}\ and\ \bibinfo {author} {\bibfnamefont {A.}~\bibnamefont
			{Guvendi}},\ }\bibfield  {title} {\bibinfo {title} {On the klein--gordon
			scalar field oscillators in a spacetime with spiral-like dislocations in
			external magnetic fields},\ }\href
	{https://doi.org/10.1140/epjc/s10052-025-13779-w} {\bibfield  {journal}
		{\bibinfo  {journal} {The European Physical Journal C}\ }\textbf {\bibinfo
			{volume} {85}},\ \bibinfo {pages} {34} (\bibinfo {year} {2025})}\BibitemShut
	{NoStop}%
	\bibitem [{\citenamefont {Zare}\ \emph {et~al.}(2022)\citenamefont {Zare},
		\citenamefont {Hassanabadi},\ and\ \citenamefont
		{Guvendi}}]{EPJP.2022.137.589}%
	\BibitemOpen
	\bibfield  {author} {\bibinfo {author} {\bibfnamefont {S.}~\bibnamefont
			{Zare}}, \bibinfo {author} {\bibfnamefont {H.}~\bibnamefont {Hassanabadi}},\
		and\ \bibinfo {author} {\bibfnamefont {A.}~\bibnamefont {Guvendi}},\
	}\bibfield  {title} {\bibinfo {title} {Relativistic landau quantization for a
			composite system in the spiral dislocation spacetime},\ }\href
	{https://doi.org/10.1140/epjp/s13360-022-02802-8} {\bibfield  {journal}
		{\bibinfo  {journal} {The European Physical Journal Plus}\ }\textbf {\bibinfo
			{volume} {137}},\ \bibinfo {pages} {589} (\bibinfo {year}
		{2022})}\BibitemShut {NoStop}%
	\bibitem [{\citenamefont {Guvendi}\ \emph {et~al.}(2021)\citenamefont
		{Guvendi}, \citenamefont {Zare},\ and\ \citenamefont
		{Hassanabadi}}]{EPJA.2021.57.192}%
	\BibitemOpen
	\bibfield  {author} {\bibinfo {author} {\bibfnamefont {A.}~\bibnamefont
			{Guvendi}}, \bibinfo {author} {\bibfnamefont {S.}~\bibnamefont {Zare}},\ and\
		\bibinfo {author} {\bibfnamefont {H.}~\bibnamefont {Hassanabadi}},\
	}\bibfield  {title} {\bibinfo {title} {Vector boson oscillator in the spiral
			dislocation spacetime},\ }\href
	{https://doi.org/10.1140/epja/s10050-021-00514-8} {\bibfield  {journal}
		{\bibinfo  {journal} {The European Physical Journal A}\ }\textbf {\bibinfo
			{volume} {57}},\ \bibinfo {pages} {192} (\bibinfo {year} {2021})}\BibitemShut
	{NoStop}%
	\bibitem [{\citenamefont {Moreira}\ and\ \citenamefont
		{Ahmed}(2024)}]{IJP.2024.98.4827}%
	\BibitemOpen
	\bibfield  {author} {\bibinfo {author} {\bibfnamefont {A.~R.~P.}\
			\bibnamefont {Moreira}}\ and\ \bibinfo {author} {\bibfnamefont
			{F.}~\bibnamefont {Ahmed}},\ }\bibfield  {title} {\bibinfo {title} {Shannon
			entropy measurements for quantum oscillator system in the presence of a
			spiral dislocation},\ }\href {https://doi.org/10.1007/s12648-024-03219-y}
	{\bibfield  {journal} {\bibinfo  {journal} {Indian Journal of Physics}\
		}\textbf {\bibinfo {volume} {98}},\ \bibinfo {pages} {4827} (\bibinfo {year}
		{2024})}\BibitemShut {NoStop}%
	\bibitem [{\citenamefont {Filgueiras}\ and\ \citenamefont
		{de~Oliveira}(2011)}]{AdP.2011.523.898}%
	\BibitemOpen
	\bibfield  {author} {\bibinfo {author} {\bibfnamefont {C.}~\bibnamefont
			{Filgueiras}}\ and\ \bibinfo {author} {\bibfnamefont {B.}~\bibnamefont
			{de~Oliveira}},\ }\bibfield  {title} {\bibinfo {title} {Electron on a
			cylinder with topological defects in a homogeneous magnetic field},\ }\href
	{https://doi.org/https://doi.org/10.1002/andp.201000158} {\bibfield
		{journal} {\bibinfo  {journal} {Annalen der Physik}\ }\textbf {\bibinfo
			{volume} {523}},\ \bibinfo {pages} {898} (\bibinfo {year}
		{2011})}\BibitemShut {NoStop}%
	\bibitem [{\citenamefont {BenDaniel}\ and\ \citenamefont
		{Duke}(1966)}]{PhysRev.152.683}%
	\BibitemOpen
	\bibfield  {author} {\bibinfo {author} {\bibfnamefont {D.~J.}\ \bibnamefont
			{BenDaniel}}\ and\ \bibinfo {author} {\bibfnamefont {C.~B.}\ \bibnamefont
			{Duke}},\ }\bibfield  {title} {\bibinfo {title} {Space-charge effects on
			electron tunneling},\ }\href {https://doi.org/10.1103/PhysRev.152.683}
	{\bibfield  {journal} {\bibinfo  {journal} {Phys. Rev.}\ }\textbf {\bibinfo
			{volume} {152}},\ \bibinfo {pages} {683} (\bibinfo {year}
		{1966})}\BibitemShut {NoStop}%
	\bibitem [{\citenamefont {von Roos}(1983)}]{vonRoos}%
	\BibitemOpen
	\bibfield  {author} {\bibinfo {author} {\bibfnamefont {O.}~\bibnamefont {von
				Roos}},\ }\bibfield  {title} {\bibinfo {title} {Position-dependent effective
			masses in semiconductor theory},\ }\href
	{https://doi.org/10.1103/PhysRevB.27.7547} {\bibfield  {journal} {\bibinfo
			{journal} {Physical Review B}\ }\textbf {\bibinfo {volume} {27}},\ \bibinfo
		{pages} {7547} (\bibinfo {year} {1983})}\BibitemShut {NoStop}%
	\bibitem [{\citenamefont {Morrow}(1987)}]{Morrow}%
	\BibitemOpen
	\bibfield  {author} {\bibinfo {author} {\bibfnamefont {R.}~\bibnamefont
			{Morrow}},\ }\bibfield  {title} {\bibinfo {title} {Position-dependent
			effective masses and band offsets},\ }\href
	{https://doi.org/10.1103/PhysRevB.35.8074} {\bibfield  {journal} {\bibinfo
			{journal} {Physical Review B}\ }\textbf {\bibinfo {volume} {35}},\ \bibinfo
		{pages} {8074} (\bibinfo {year} {1987})}\BibitemShut {NoStop}%
	\bibitem [{\citenamefont {Alhaidari}(2002)}]{PhysRevA.66.042116}%
	\BibitemOpen
	\bibfield  {author} {\bibinfo {author} {\bibfnamefont {A.~D.}\ \bibnamefont
			{Alhaidari}},\ }\bibfield  {title} {\bibinfo {title} {Solutions of the
			nonrelativistic wave equation with position-dependent effective mass},\
	}\href {https://doi.org/10.1103/PhysRevA.66.042116} {\bibfield  {journal}
		{\bibinfo  {journal} {Phys. Rev. A}\ }\textbf {\bibinfo {volume} {66}},\
		\bibinfo {pages} {042116} (\bibinfo {year} {2002})}\BibitemShut {NoStop}%
	\bibitem [{\citenamefont {Plastino}\ \emph {et~al.}(1999)\citenamefont
		{Plastino}, \citenamefont {Rigo}, \citenamefont {Casas}, \citenamefont
		{Garcias},\ and\ \citenamefont {Plastino}}]{PRA.1999.60.4318}%
	\BibitemOpen
	\bibfield  {author} {\bibinfo {author} {\bibfnamefont {A.~R.}\ \bibnamefont
			{Plastino}}, \bibinfo {author} {\bibfnamefont {A.}~\bibnamefont {Rigo}},
		\bibinfo {author} {\bibfnamefont {M.}~\bibnamefont {Casas}}, \bibinfo
		{author} {\bibfnamefont {F.}~\bibnamefont {Garcias}},\ and\ \bibinfo {author}
		{\bibfnamefont {A.}~\bibnamefont {Plastino}},\ }\bibfield  {title} {\bibinfo
		{title} {Supersymmetric approach to quantum systems with position-dependent
			effective mass},\ }\href {https://doi.org/10.1103/PhysRevA.60.4318}
	{\bibfield  {journal} {\bibinfo  {journal} {Phys. Rev. A}\ }\textbf {\bibinfo
			{volume} {60}},\ \bibinfo {pages} {4318} (\bibinfo {year}
		{1999})}\BibitemShut {NoStop}%
	\bibitem [{\citenamefont {DeWitt}(1957)}]{dewitt1957dynamical}%
	\BibitemOpen
	\bibfield  {author} {\bibinfo {author} {\bibfnamefont {B.~S.}\ \bibnamefont
			{DeWitt}},\ }\bibfield  {title} {\bibinfo {title} {Dynamical theory in curved
			spaces. i. a review of the classical and quantum action principles},\ }\href
	{https://doi.org/10.1103/RevModPhys.29.377} {\bibfield  {journal} {\bibinfo
			{journal} {Rev. Mod. Phys.}\ }\textbf {\bibinfo {volume} {29}},\ \bibinfo
		{pages} {377} (\bibinfo {year} {1957})}\BibitemShut {NoStop}%
	\bibitem [{\citenamefont {R.{ }C.{ }T.{ }da{ }Costa}(1981)}]{dacosta}%
	\BibitemOpen
	\bibfield  {author} {\bibinfo {author} {\bibnamefont {R.{ }C.{ }T.{ }da{
				}Costa}},\ }\bibfield  {title} {\bibinfo {title} {Quantum mechanics of a
			constrained particle},\ }\href {https://doi.org/10.1103/PhysRevA.23.1982}
	{\bibfield  {journal} {\bibinfo  {journal} {Phys. Rev. A}\ }\textbf {\bibinfo
			{volume} {23}},\ \bibinfo {pages} {1982} (\bibinfo {year}
		{1981})}\BibitemShut {NoStop}%
	\bibitem [{\citenamefont {Tan}\ and\ \citenamefont
		{Inkson}(1996)}]{TanInkson1996}%
	\BibitemOpen
	\bibfield  {author} {\bibinfo {author} {\bibfnamefont {W.-C.}\ \bibnamefont
			{Tan}}\ and\ \bibinfo {author} {\bibfnamefont {J.~C.}\ \bibnamefont
			{Inkson}},\ }\bibfield  {title} {\bibinfo {title} {Electron states in a
			two-dimensional ring---an exactly soluble model},\ }\href
	{https://doi.org/10.1088/0268-1242/11/11/001} {\bibfield  {journal} {\bibinfo
			{journal} {Semiconductor Science and Technology}\ }\textbf {\bibinfo
			{volume} {11}},\ \bibinfo {pages} {1635} (\bibinfo {year}
		{1996})}\BibitemShut {NoStop}%
	\bibitem [{\citenamefont {Rezaei}\ \emph {et~al.}(2011)\citenamefont {Rezaei},
		\citenamefont {Vahdani},\ and\ \citenamefont {Vaseghi}}]{CAP.2011.11.181}%
	\BibitemOpen
	\bibfield  {author} {\bibinfo {author} {\bibfnamefont {G.}~\bibnamefont
			{Rezaei}}, \bibinfo {author} {\bibfnamefont {M.}~\bibnamefont {Vahdani}},\
		and\ \bibinfo {author} {\bibfnamefont {B.}~\bibnamefont {Vaseghi}},\
	}\bibfield  {title} {\bibinfo {title} {Nonlinear optical properties of a
			hydrogenic impurity in an ellipsoidal finite potential quantum dot},\ }\href
	{https://doi.org/https://doi.org/10.1016/j.cap.2010.07.002} {\bibfield
		{journal} {\bibinfo  {journal} {Current Applied Physics}\ }\textbf {\bibinfo
			{volume} {11}},\ \bibinfo {pages} {176} (\bibinfo {year} {2011})}\BibitemShut
	{NoStop}%
	\bibitem [{\citenamefont {Bahar}\ and\ \citenamefont
		{Başer}(2023)}]{PBCM.2023.665.415042}%
	\BibitemOpen
	\bibfield  {author} {\bibinfo {author} {\bibfnamefont {M.~K.}\ \bibnamefont
			{Bahar}}\ and\ \bibinfo {author} {\bibfnamefont {P.}~\bibnamefont {Başer}},\
	}\bibfield  {title} {\bibinfo {title} {The second, third harmonic generations
			and nonlinear optical rectification of the mathieu quantum dot with the
			external electric, magnetic and laser field},\ }\href
	{https://doi.org/https://doi.org/10.1016/j.physb.2023.415042} {\bibfield
		{journal} {\bibinfo  {journal} {Physica B: Condensed Matter}\ }\textbf
		{\bibinfo {volume} {665}},\ \bibinfo {pages} {415042} (\bibinfo {year}
		{2023})}\BibitemShut {NoStop}%
\end{thebibliography}
\end{document}